\documentclass[
aps,
prd,
twocolumn,
showpacs,
preprintnumbers,
amsmath,
amssymb,
superscriptaddress,
altaffilletter,
reprint]
{revtex4-1}

\usepackage{relsize}
\usepackage{longtable}
\usepackage{graphicx}
\usepackage{hyperref}
\usepackage{epsfig}
\usepackage{amssymb}
\usepackage{latexsym} 
\usepackage{multirow}
\usepackage{color}
\usepackage[usenames,dvipsnames,svgnames,table]{xcolor}

\begin{document}
\title{Measurement of the inclusive $\nu_{\mu}$ charged current cross section on carbon in the near detector of the T2K experiment}

\date{\today}


\newcommand{\INSTC}{\affiliation{University of Alberta, Centre for Particle Physics, Department of Physics, Edmonton, Alberta, Canada}}
\newcommand{\INSTEE}{\affiliation{University of Bern, Albert Einstein Center for Fundamental Physics, Laboratory for High Energy Physics (LHEP), Bern, Switzerland}}
\newcommand{\INSTFE}{\affiliation{Boston University, Department of Physics, Boston, Massachusetts, U.S.A.}}
\newcommand{\INSTD}{\affiliation{University of British Columbia, Department of Physics and Astronomy, Vancouver, British Columbia, Canada}}
\newcommand{\INSTGA}{\affiliation{University of California, Irvine, Department of Physics and Astronomy, Irvine, California, U.S.A.}}
\newcommand{\INSTI}{\affiliation{IRFU, CEA Saclay, Gif-sur-Yvette, France}}
\newcommand{\INSTCI}{\affiliation{Chonnam National University, Institute for Universe \& Elementary Particles, Gwangju, Korea}}
\newcommand{\INSTGB}{\affiliation{University of Colorado at Boulder, Department of Physics, Boulder, Colorado, U.S.A.}}
\newcommand{\INSTFG}{\affiliation{Colorado State University, Department of Physics, Fort Collins, Colorado, U.S.A.}}
\newcommand{\INSTCJ}{\affiliation{Dongshin University, Department of Physics, Naju, Korea}}
\newcommand{\INSTFH}{\affiliation{Duke University, Department of Physics, Durham, North Carolina, U.S.A.}}
\newcommand{\INSTBA}{\affiliation{Ecole Polytechnique, IN2P3-CNRS, Laboratoire Leprince-Ringuet, Palaiseau, France }}
\newcommand{\INSTEF}{\affiliation{ETH Zurich, Institute for Particle Physics, Zurich, Switzerland}}
\newcommand{\INSTEG}{\affiliation{University of Geneva, Section de Physique, DPNC, Geneva, Switzerland}}
\newcommand{\INSTDG}{\affiliation{H. Niewodniczanski Institute of Nuclear Physics PAN, Cracow, Poland}}
\newcommand{\INSTCB}{\affiliation{High Energy Accelerator Research Organization (KEK), Tsukuba, Ibaraki, Japan}}
\newcommand{\INSTED}{\affiliation{Institut de Fisica d'Altes Energies (IFAE), Bellaterra (Barcelona), Spain}}
\newcommand{\INSTEC}{\affiliation{IFIC (CSIC \& University of Valencia), Valencia, Spain}}
\newcommand{\INSTEI}{\affiliation{Imperial College London, Department of Physics, London, United Kingdom}}
\newcommand{\INSTGF}{\affiliation{INFN Sezione di Bari and Universit\`a e Politecnico di Bari, Dipartimento Interuniversitario di Fisica, Bari, Italy}}
\newcommand{\INSTBE}{\affiliation{INFN Sezione di Napoli and Universit\`a di Napoli, Dipartimento di Fisica, Napoli, Italy}}
\newcommand{\INSTBF}{\affiliation{INFN Sezione di Padova and Universit\`a di Padova, Dipartimento di Fisica, Padova, Italy}}
\newcommand{\INSTBD}{\affiliation{INFN Sezione di Roma and Universit\`a di Roma ``La Sapienza'', Roma, Italy}}
\newcommand{\INSTEB}{\affiliation{Institute for Nuclear Research of the Russian Academy of Sciences, Moscow, Russia}}
\newcommand{\INSTCC}{\affiliation{Kobe University, Kobe, Japan}}
\newcommand{\INSTCD}{\affiliation{Kyoto University, Department of Physics, Kyoto, Japan}}
\newcommand{\INSTEJ}{\affiliation{Lancaster University, Physics Department, Lancaster, United Kingdom}}
\newcommand{\INSTFC}{\affiliation{University of Liverpool, Department of Physics, Liverpool, United Kingdom}}
\newcommand{\INSTFI}{\affiliation{Louisiana State University, Department of Physics and Astronomy, Baton Rouge, Louisiana, U.S.A.}}
\newcommand{\INSTJ}{\affiliation{Universit\'e de Lyon, Universit\'e Claude Bernard Lyon 1, IPN Lyon (IN2P3), Villeurbanne, France}}
\newcommand{\INSTCE}{\affiliation{Miyagi University of Education, Department of Physics, Sendai, Japan}}
\newcommand{\INSTDF}{\affiliation{National Centre for Nuclear Research, Warsaw, Poland}}
\newcommand{\INSTFJ}{\affiliation{State University of New York at Stony Brook, Department of Physics and Astronomy, Stony Brook, New York, U.S.A.}}
\newcommand{\INSTCF}{\affiliation{Osaka City University, Department of Physics, Osaka,  Japan}}
\newcommand{\INSTGG}{\affiliation{Oxford University, Department of Physics, Oxford, United Kingdom}}
\newcommand{\INSTBB}{\affiliation{UPMC, Universit\'e Paris Diderot, CNRS/IN2P3, Laboratoire de Physique Nucl\'eaire et de Hautes Energies (LPNHE), Paris, France}}
\newcommand{\INSTGC}{\affiliation{University of Pittsburgh, Department of Physics and Astronomy, Pittsburgh, Pennsylvania, U.S.A.}}
\newcommand{\INSTFA}{\affiliation{Queen Mary, University of London, School of Physics and Astronomy, London, United Kingdom}}
\newcommand{\INSTE}{\affiliation{University of Regina, Department of Physics, Regina, Saskatchewan, Canada}}
\newcommand{\INSTGD}{\affiliation{University of Rochester, Department of Physics and Astronomy, Rochester, New York, U.S.A.}}
\newcommand{\INSTBC}{\affiliation{RWTH Aachen University, III. Physikalisches Institut, Aachen, Germany}}
\newcommand{\INSTDD}{\affiliation{Seoul National University, Department of Physics and Astronomy, Seoul, Korea}}
\newcommand{\INSTFB}{\affiliation{University of Sheffield, Department of Physics and Astronomy, Sheffield, United Kingdom}}
\newcommand{\INSTDI}{\affiliation{University of Silesia, Institute of Physics, Katowice, Poland}}
\newcommand{\INSTEH}{\affiliation{STFC, Rutherford Appleton Laboratory, Harwell Oxford,  and  Daresbury Laboratory, Warrington, United Kingdom}}
\newcommand{\INSTCH}{\affiliation{University of Tokyo, Department of Physics, Tokyo, Japan}}
\newcommand{\INSTBJ}{\affiliation{University of Tokyo, Institute for Cosmic Ray Research, Kamioka Observatory, Kamioka, Japan}}
\newcommand{\INSTCG}{\affiliation{University of Tokyo, Institute for Cosmic Ray Research, Research Center for Cosmic Neutrinos, Kashiwa, Japan}}
\newcommand{\INSTF}{\affiliation{University of Toronto, Department of Physics, Toronto, Ontario, Canada}}
\newcommand{\INSTB}{\affiliation{TRIUMF, Vancouver, British Columbia, Canada}}
\newcommand{\INSTG}{\affiliation{University of Victoria, Department of Physics and Astronomy, Victoria, British Columbia, Canada}}
\newcommand{\INSTDJ}{\affiliation{University of Warsaw, Faculty of Physics, Warsaw, Poland}}
\newcommand{\INSTDH}{\affiliation{Warsaw University of Technology, Institute of Radioelectronics, Warsaw, Poland}}
\newcommand{\INSTFD}{\affiliation{University of Warwick, Department of Physics, Coventry, United Kingdom}}
\newcommand{\INSTGE}{\affiliation{University of Washington, Department of Physics, Seattle, Washington, U.S.A.}}
\newcommand{\INSTGH}{\affiliation{University of Winnipeg, Department of Physics, Winnipeg, Manitoba, Canada}}
\newcommand{\INSTEA}{\affiliation{Wroclaw University, Faculty of Physics and Astronomy, Wroclaw, Poland}}
\newcommand{\INSTH}{\affiliation{York University, Department of Physics and Astronomy, Toronto, Ontario, Canada}}

\INSTC
\INSTEE
\INSTFE
\INSTD
\INSTGA
\INSTI
\INSTCI
\INSTGB
\INSTFG
\INSTCJ
\INSTFH
\INSTBA
\INSTEF
\INSTEG
\INSTDG
\INSTCB
\INSTED
\INSTEC
\INSTEI
\INSTGF
\INSTBE
\INSTBF
\INSTBD
\INSTEB
\INSTCC
\INSTCD
\INSTEJ
\INSTFC
\INSTFI
\INSTJ
\INSTCE
\INSTDF
\INSTFJ
\INSTCF
\INSTGG
\INSTBB
\INSTGC
\INSTFA
\INSTE
\INSTGD
\INSTBC
\INSTDD
\INSTFB
\INSTDI
\INSTEH
\INSTCH
\INSTBJ
\INSTCG
\INSTF
\INSTB
\INSTG
\INSTDJ
\INSTDH
\INSTFD
\INSTGE
\INSTGH
\INSTEA
\INSTH

\author{K.\,Abe}\INSTBJ
\author{N.\,Abgrall}\INSTEG
\author{H.\,Aihara}\thanks{also at Kavli IPMU, U. of Tokyo, Kashiwa, Japan}\INSTCH
\author{T.\,Akiri}\INSTFH
\author{J.B.\,Albert}\INSTFH
\author{C.\,Andreopoulos}\INSTEH
\author{S.\,Aoki}\INSTCC
\author{A.\,Ariga}\INSTEE
\author{T.\,Ariga}\INSTEE
\author{S.\,Assylbekov}\INSTFG
% ...

\collaboration{The T2K Collaboration}\noaffiliation

\pagebreak
\begin{abstract}
T2K has performed the first measurement of  $\nu_{\mu}$ inclusive charged current interactions on carbon at neutrino energies of $\sim$1~GeV where the measurement is reported as a flux-averaged double differential cross section in muon momentum and angle. 
The flux is predicted by the beam Monte Carlo and external data, including the results from the NA61/SHINE experiment. The data used for this measurement were taken in 2010 and 2011, with a total of $10.8 \times 10^{19}$ protons-on-target. The analysis is performed on 4485 inclusive charged current interaction candidates selected in the most upstream fine-grained scintillator detector of the near detector. 
 The flux-averaged total cross section is $\langle \sigma_{ \rm{CC}} \rangle_{\phi} =(6.91 \pm 0.13 (stat) \pm 0.84 (syst)) \times 10^{-39} \rm{\frac{cm^2}{nucleon}}$ for a mean neutrino energy of 0.85~GeV.
\end{abstract}

\maketitle

\section{Introduction}

The T2K (Tokai-to-Kamioka) experiment is a long baseline neutrino oscillation experiment \cite{Abe:2011ks} whose primary goals are to make precise measurements of the appearance of electron neutrinos and the disappearance of muon neutrinos at a distance where the oscillation is maximal for the neutrino beam energy. This analysis is motivated, in part, by the fact that improved precision in the oscillation analyses requires better knowledge of neutrino interaction cross sections.

Charged current (CC) neutrino-nucleon interactions at neutrino energies around 1 GeV have been studied in the past predominantly on deuterium targets \cite{PhysRevD.19.2521,PhysRevD.25.617}.
Many modern neutrino oscillation experiments like SciBooNE, MiniBooNE and T2K use heavier targets rich in carbon (and/or oxygen). Nuclear effects are important for those targets and, consequently, the neutrino interaction cross sections are not well known. It is therefore important to measure and understand these interactions to minimize systematic uncertainties for the oscillation measurement.

For this purpose, we present a flux-averaged double differential inclusive muon neutrino CC cross section on carbon as a function of muon momentum and angle.
An unfolding method that corrects for detector resolution, efficiencies and backgrounds is used to extract the flux-averaged total cross section and differential cross section in muon momentum and angle. The cross-section extraction technique requires a detailed Monte Carlo (MC) simulation including the neutrino flux prediction (Sec.~\ref{sec:flux}), a complete neutrino interaction model (Sec.~\ref{sec:nuint}), full description of the near detector and accurate simulation of the readout electronics. The neutrino event generator NEUT \cite{Hayato:2009} is used in the calculations of the selection efficiency for signal events and the rate of background events in the analysis. Additionally, a different event generator, GENIE \cite{Andreopoulos:2009rq}, is used for cross-checks and for fake data studies. Finally, the results of this analysis are compared with the predictions of the NEUT and GENIE generators.

In this paper, we first summarize the T2K experiment in Sec.~\ref{sec:T2Kexp}, where we present a description of the off-axis near detector (Sec.~\ref{sec:nd280oa}) and, in particular, the tracker region where the neutrino interaction target is located. Section~\ref{sec:nuint} is devoted to a detailed description of how neutrino interactions are simulated in T2K. Section~\ref{sec:ccsel} summarizes the reconstruction tools and the event selection in this analysis. The method used to extract the cross section is explained in Sec.~\ref{sec:ccxs} and the results are presented in Sec.~\ref{sec:results}. The conclusions are given in Sec.~\ref{sec:conclusion}.

\section{T2K experiment}\label{sec:T2Kexp}
T2K consists of an accelerator-generated neutrino beamline, a near detector complex 280~m downstream of the neutrino beam target and a far detector, Super-Kamiokande (SK),
located 295 km away at an angle of 2.5 degrees from the axis of the neutrino beam. 
Neutrinos are generated from the 30 GeV proton beam of the Japan Proton Accelerator Research Complex (J-PARC) located in Tokai-mura on the east coast of Japan.
The near detector complex is composed of a detector on the axis of the neutrino beam, called INGRID, and a detector located 2.5 degrees off axis, in line with SK, called ND280. INGRID is used primarily to measure the beam profile and stability, and the ND280 detector is used to measure neutrino fluxes and neutrino interaction cross-section properties, as the inclusive CC cross section presented in this paper. 

\begin{figure*}[htbp]
\begin{center}
  \includegraphics[width=1\textwidth]{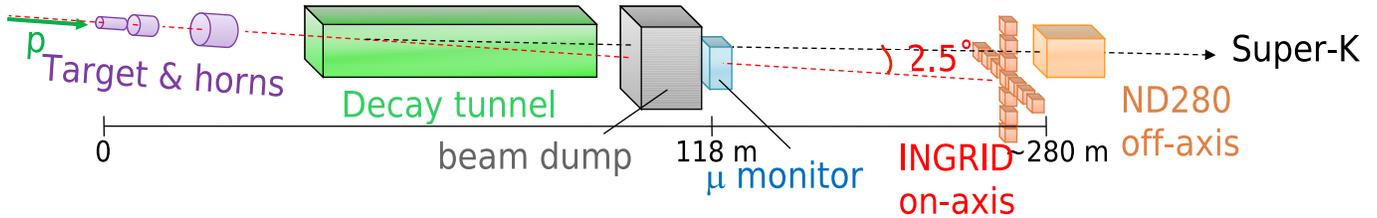}
\caption{A schematic view of the T2K neutrino beamline and near detectors.}
\label{fig:t2kndsk}
\end{center}
\end{figure*}

In the following three subsections, the experiment is discussed in detail.
We provide an overview of the neutrino beamline in Sec.~\ref{sec:beamline}. 
Section~\ref{sec:flux} describes the T2K flux simulation included in the MCs used to extract the cross section, which represents a major systematic uncertainty.
Section~\ref{sec:nd280oa} describes the off-axis detector, emphasizing the tracking detectors and target composition, as these are essential for the cross-section calculation. 
 We refer to \cite{Abe:2011ks} for a complete description of the T2K experiment.

\subsection{Neutrino beam}\label{sec:beamline}
Figure~\ref{fig:t2kndsk} depicts the neutrino beamline and the near detectors.
The neutrino beam is produced by protons accelerated to 30 GeV kinetic energy.
The proton beam has eight bunches (six before June 2010) with a 581~ns spacing. 
 The protons in the spill are extracted and directed toward a 91.4 cm long graphite target aligned $2.5^\circ$ off-axis angle from Kamioka.
The target is installed inside a magnetic horn that collects and focuses the positively charged mesons (mainly pions and kaons) generated by proton interactions in the target. 
Two additional magnetic horns are used to further focus the charged mesons before they enter a 96 m long steel decay volume filled with helium. The mesons decay predominantly into highly boosted muons and muon neutrinos, which propagate roughly in the direction of the decaying mesons.
 A beam dump stops most of the particles in the beam that are not neutrinos. Some high-energy muons pass through the beam dump and are observed by the muon monitor, providing information used to track the beam direction and stability.
 The analysis presented in this paper uses the data taken before March 2011, comprising a total of $10.8\times 10^{19}$ protons-on-target (POT).

\subsection{Neutrino flux prediction} \label{sec:flux}
\begin{figure}[htbp]
\begin{center}
 \includegraphics[width=0.49\textwidth]{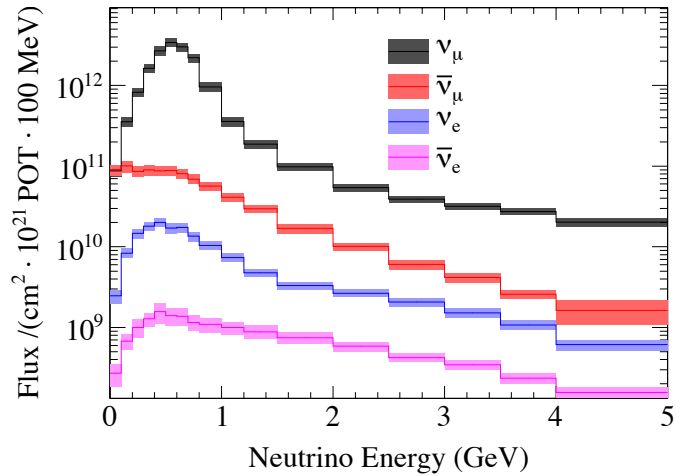}
\caption{ The ND280 flux prediction with systematic error bars, for each neutrino species.
The prediction takes into account the correct relative fractions of 2010 and 2011 beam conditions.}
\label{fig:nuflux_beam}
\end{center}
\end{figure}

A detailed description of the neutrino flux prediction can be found in \cite{Abe:2012av}.
A FLUKA2008 \cite{Battistoni:2007zzb,Ferrari:2005zk} and
GEANT3.21/GCALOR \cite{Brun:1994,Zeitnitz:1994bs} based simulation models
the physical processes involved in the neutrino production, from the interaction of primary beam protons in the T2K target, to the decay of hadrons and muons that produce neutrinos.
The simulation uses T2K proton beam monitor measurements as inputs.
The modeling of hadronic interactions is reweighted using thin target hadron production data, including recent charged pion and kaon measurements from the NA61/SHINE experiment~\cite{Abgrall:2011ae,Abgrall:2011ts}, which cover most of the kinematic region of interest. 
The predicted neutrino fluxes and energy spectra at the near detector are shown in Fig.~\ref{fig:nuflux_beam}.
The integrated muon neutrino flux in the chosen fiducial volume for our data exposure is 2.02$\times 10^{12}$ cm$^{-2}$.

For the first published T2K oscillation analyses, the uncertainty on the predicted neutrino flux for this beam was as large as 20\% \cite{Abe:2011sj,Abe:2012gx}. For this work, however, the tuning of the particle production model to NA61/SHINE measurements led to a substantial reduction in systematic errors in the flux.
With the latest results released by the NA61/SHINE collaboration on the kaon production cross section based on the 2007 data \cite{Abgrall:2011ts}, the uncertainty of the integrated flux is now about 11\%.

The parametrization of the flux uncertainties is described by normalization parameters in bins of neutrino energy and flavor at the near detector. The different sources of uncertainty can be separated into two categories: the hadron production uncertainty and the T2K beamline uncertainty.

The uncertainties on hadron production are mainly driven by the NA61/SHINE measurements and the Eichten and Allaby data \cite{Eichten:1972nw,Allaby:1970jt}, and constitute the dominant component of the flux uncertainty. They include the uncertainty in the production cross section, the secondary nucleon production, pion production multiplicity and kaon production multiplicity.

The second category of flux uncertainties is associated with operational variations in the beamline conditions during the data taking. They include uncertainties in the proton beam positioning, the off-axis angle, the horn absolute current, the horn angular alignment, the horn field asymmetry, the target alignment, the position dependence of the flux in the near detector and the proton beam intensity. The last two uncertainties were found to be very small and are therefore considered negligible.
 Table \ref{tab:flux_summary} shows the contribution of each source of uncertainty to the total uncertainty.

\begin{table}[!htbp]
\caption{\label{tab:flux_summary} The contribution of each source to the total muon neutrino flux uncertainty.}
\centering
\begin{tabular}{c| c }
\hline
\hline
 Error source              &  Error (\%) \\
\hline
Production cross section     & 6.4  \\  
Secondary nucleon production       & 6.9  \\ 
Pion multiplicity          & 5.0  \\  
Kaon multiplicity          & 0.8  \\    
Off-axis angle             & 1.6  \\  
Proton beam                & 1.1  \\  
Horn absolute current          & 0.9  \\ 
Horn angular alignment           & 0.5  \\  
Horn field asymmetry            & 0.3  \\  
Target alignment               & 0.2  \\   
\hline
Total                   & 10.9  \\
\hline
\hline
\end{tabular}
\end{table}

\subsection{The off-axis ND280 detector}\label{sec:nd280oa}

\begin{figure}[tbh]
  \begin{center}
    \includegraphics[width=0.85\linewidth]{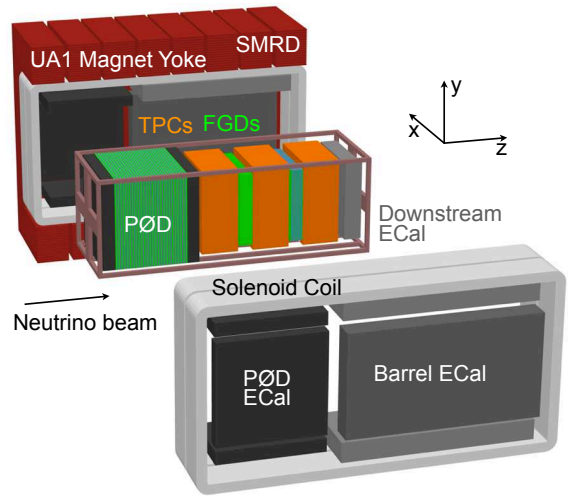}
  \end{center}
  \caption{An exploded view of the ND280 off-axis detector.}
  \label{fig:nd280-exploded}
\end{figure}

The ND280 detector is a magnetized particle tracking device. The active elements are contained inside a large magnet, which was previously used for the UA1 and NOMAD experiments at CERN. Inside the upstream end of this magnet sits a $\pi^0$ detector (P\O{}D) consisting of tracking planes of scintillating bars alternating with either a water target and a brass foil, or a lead foil. Downstream of the P\O{}D, the tracker consists of three gas-filled time projection chambers (TPCs) \cite{T2KtpcNIM2010} and two fine-grained detectors (FGDs) \cite{Amaudruz:2012pe} made up of finely segmented scintillating bars. The tracker is designed to measure neutrino interactions in the FGDs. The P\O{}D, TPCs, and FGDs are all surrounded by an electromagnetic calorimeter (ECal) which detects $\gamma$ rays that fail to convert in the inner detectors, while the return yoke of the magnet is instrumented with a scintillator to aid in the identification and range determination for muons that exit out the sides of the off-axis detector. Figure~\ref{fig:nd280-exploded} shows an exploded view of ND280.

 The main detectors used for the analysis presented in this paper are the FGDs and TPCs.
For this measurement, the most upstream FGD (FGD1) is used as an active target for the neutrino interactions, while the second FGD (FGD2) and the TPCs act as tracking detectors.

The FGDs are constructed from 9.61~mm
$\times$ 9.61~mm $\times$ 1864.3~mm bars of extruded polystyrene
scintillator, which are oriented along two orthogonal directions perpendicular to the beam. The ND280 coordinate system is a right-handed Cartesian coordinate system with the $z$ axis running through the central axis of the tracker in the beam direction and the $y$ axis in the upward vertical direction.

\begin{figure}[hbpt]
  \centering
  \includegraphics[width=0.5\textwidth]{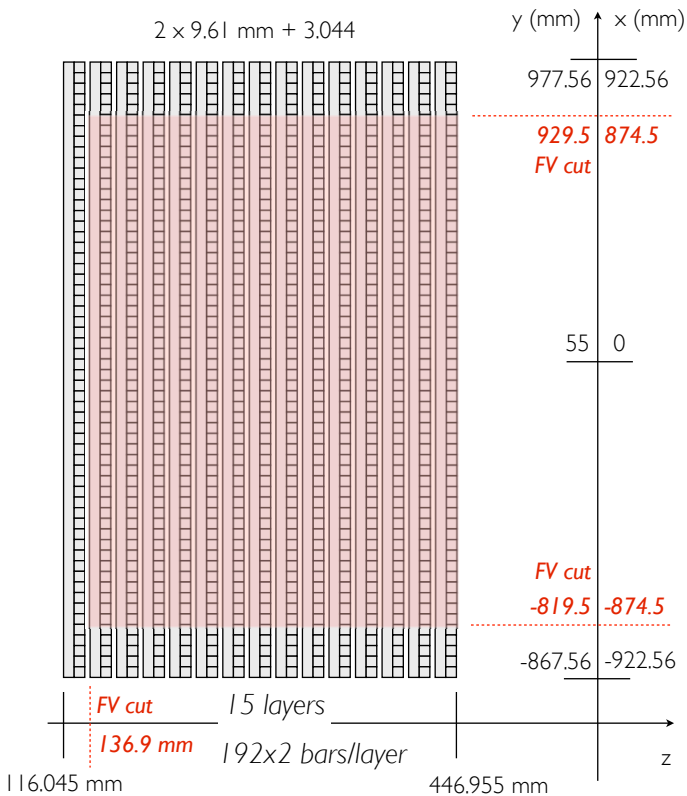}
  \caption{Schematic view of FGD1 and its fiducial volume (FV) delimited by the dashed line. In the $y$ coordinate, the detector is displaced 55 mm upwards relative to the center of the ND280 coordinate system, which is centered in the magnet.}
  \label{fig:fgd}
\end{figure}

The FGD1 consists of 5760
scintillator bars, arranged in 30 layers of 192 bars each, with the layers
oriented alternately in the $x$ and $y$ directions. The alternating directions of the bars allow three-dimensional tracking of the charged particles.
The bars are glued to thin sheets of a glass fiber laminate (G-10) and to each other using Plexus MA590 adhesive. The scintillator bars consist of polystyrene doped with PPO (2,5-Diphenyloxazole) and POPOP (1,4-bis(5-phenyloxazol-2-yl) benzene) plus a thin reflective coextruded TiO$_2$ coating. A wavelength shifting fiber, read out by a MPPC (multi-pixel photon counter), is embedded inside each bar.
The composition of the FGD is carbon (86.0\%), hydrogen (7.4\%), oxygen (${3.7\%}$), titanium (${1.7\%}$), silicon (${1.0\%}$) and nitrogen (${0.1\%}$), where the percentages represent the mass fraction of each element.
The number of nucleons is given by
\begin{eqnarray}
T&=&   N_{A}  V_{FV} \cdot \rho_{scint}\cdot\displaystyle\sum_{\rm{a=C,O,H,Ti,Si,N}} f_{a} \frac{A_{a}}{M_{a}}\label{eq:Tnucleon}\\
&=&5.50\times 10^{29} \rm{nucleons}\nonumber
\end{eqnarray}
where $N_A=6.022 \times 10^{23}$~mol$^{-1}$ is Avogadro's number; $V_{FV}$ is the fiducial volume (FV) inside FGD1 (see Fig.~\ref{fig:fgd}); $\rho_{scint}= 0.963~\rm{g/cm}^3$ is the density of the scintillator inside the fiducial volume including the glue, coating and the air gaps between the modules; $a$ runs over the elements present in the scintillator; $f_a$ is the mass fraction; $A_{a}$ represents the averaged number of nucleons per nucleus; and $M_a$ is the atomic mass.
The ratio of protons to neutrons is 53.6:46.4.

In this analysis, the fiducial volume in the beam direction begins at the second $x$-$y$ module (as encountered by the beam), leaving the first $x$-$y$ module as a veto for incoming particles from upstream interactions. The downstream end of the fiducial volume includes the last layer of the scintillator. In the $x$ and $y$ directions, a distance equivalent to five bars on each side of the FGD lies outside the fiducial volume. Again, this provides a veto for particles generated by interactions outside the FV. 

The second FGD is a water-rich detector consisting of seven $x$-$y$ modules
of plastic scintillator alternating with six 2.5~cm
thick layers of water. It is used as a tracking detector in this analysis.

The two FGDs are sandwiched amongst three TPCs.
Each TPC consists of an inner box, which is filled with an Ar:CF$_4$:iC$_4$H$_{10}$ (95:3:2 by mass) drift gas mixture, 
within an outer box containing CO$_2$ acting as an insulating gas.
The inner (outer) walls are made from composite panels with copper-clad G-10
(aluminum) skins.
The inner box panels were precisely machined with an 11.5~mm pitch copper
strip pattern which, in conjunction with a
central cathode panel, produces a uniform electric drift field in the
active drift volume of the TPC, roughly aligned with the
magnetic field provided by the magnet.

At each side of the TPCs, 12 micromegas (micro-mesh gas detector) modules are arranged in two vertical columns.
The role of the micromegas is to amplify the ionization signals created by charged particles passing through the TPCs.

The micromegas anode is segmented into pads of 7.0~mm $\times$ 9.8~mm (vertical $\times$ horizontal) allowing 3D track reconstruction of charged particles traversing the TPC. 
The $x$ coordinate is obtained by projecting the arrival times of the pad signals as discussed in Sec. IV.B.

The TPCs perform three key functions in the near detector: three-dimensional reconstruction of charged particles crossing the detector; measurement of the momenta of charged particles via curvature in a 0.2~T magnetic field; and particle identification (PID) for charged particles via ionization (dE/dx).
These three functions allow the selection of high purity samples of different types of neutrino interactions.


\section{Simulation of neutrino interactions}\label{sec:nuint}
Neutrino interactions in the entire detector are simulated with the NEUT~\cite{Hayato:2009} and GENIE~\cite{Andreopoulos:2009rq} program libraries.
While NEUT is used as the primary generator, including for the propagation of systematic errors, GENIE is used for cross-checks and for the generation of fake data sets. NEUT and GENIE cover a similar range of neutrino energies from several tens of MeV to hundreds of TeV and are able to simulate all the nuclear targets present in the near detector.

The neutrino interactions are simulated in both neutral current (NC) and CC modes and include the processes of elastic scattering, quasielastic scattering, single meson production, single gamma production, coherent pion production and nonresonant inelastic scattering. Interactions of hadrons produced inside the nuclear medium are also simulated.

Simulations are a critical aspect of this analysis. 
In the simulated data, neutrino interactions are generated outside and within the full detector volume including all active and inactive material. This provides information necessary to understand the signal as well as backgrounds from interactions outside the fiducial volume.
The details of the simulation process are presented in the remainder of this section. 
We first describe, in Sec.~\ref{sec:numodel}, the various interaction models simulated in the event generators. 
The modeling is important to understand the selection efficiency and backgrounds, as well as how reconstructed and true quantities are related.
 In Sec.~\ref{sec:modelsyst}, we present the systematic uncertainties associated with each kind of interaction. In Sec.~\ref{sec:ccxs}, these uncertainties are propagated to the reported cross section.

\subsection{Neutrino interaction model}\label{sec:numodel}
In the following, the main interactions related to our CC-inclusive measurement are discussed. Table~\ref{tab:mode} shows the relative fractions of each neutrino interaction type simulated in our mixed target. The Monte Carlo predicts that 89\% of interactions occur on carbon.
 Some specific interactions such as NC elastic scattering, charm production and neutrino-electron elastic scattering are not discussed in this paper due to their irrelevance to this analysis. We refer to~\cite{Hayato:2009,Andreopoulos:2009rq} for the details.

\begin{table}[htpb]
\caption{\label{tab:mode} Fraction of neutrino interaction processes simulated in the total FGD1 volume by NEUT and GENIE. }
\centering
\begin{tabular}{l | c  c }
\hline
\hline
 Channel & NEUT (\%) & GENIE (\%)\\
\hline
CC quasielastic & 39.4 & 37.0
\\
CC 1$\pi$ resonant production & 18.9 & 19.9
\\
CC coherent $\pi$ & 1.6 & 0.5
\\
CC other & 11.8 & 13.8
\\
NC & 28.4 & 28.7\\
\hline
\hline
\end{tabular}
\end{table}

\subsubsection{quasielastic scattering}
Both generators use the Llewellyn Smith formalism~\cite{LlewellynSmith:1971zm} for quasielastic (QE) scattering.
In this model the hadronic weak current is expressed in terms of the most general Lorentz-invariant form factors which do not violate G-parity: namely two vector form factors, one pseudoscalar form factor and one axial form factor.
The vector form factors are measured over a broad range of kinematics in electron elastic scattering experiments and are fixed by the conserved vector current (CVC) hypothesis. The pseudoscalar form factor is assumed to have the form suggested by the partially conserved axial current (PCAC) hypothesis~\cite{LlewellynSmith:1971zm}. This leaves the axial form factor $F_A(Q^2)$ as the sole remaining unknown quantity. The value of the axial form factor at $Q^2=0$, $F_A(0)$, is well known from measurements of neutron beta decay and the $Q^2$ dependence of this form factor can only be determined in neutrino experiments.
 In both generators, a dipole form factor is assumed with an axial mass, $M_A^{QE}$, of 0.99 GeV/c$^2$ for GENIE and 1.21 GeV/c$^2$ for NEUT.

NEUT and GENIE use slightly different approaches in the treatment of nuclear effects. Both NEUT and GENIE use the relativistic Fermi gas model (RFG) to describe the nuclear environment. The RFG assumes that nucleons are bound in a potential well of binding energy, $E_B$, and that they are quasifree with a maximum momentum set to the momentum at the Fermi sphere, $p_F$. The values of $p_F$ and $E_B$ are measured from electron scattering for each target nucleus species individually.
While NEUT uses the analytical expressions from the Smith-Moniz~\cite{Smith:1972xh} model, GENIE incorporates short range nucleon-nucleon correlations in the RFG model
and handles kinematics for off-shell scattering according to the model of Bodek and Ritchie~\cite{Bodek:1981wr}.

\subsubsection{Meson production via baryon resonances}
NEUT and GENIE use the Rein-Sehgal~\cite{Rein:1980wg} model to simulate neutrino-induced single pion production.
Of the 18 resonances in the original Rein-Sehgal paper, the 16 listed as unambiguous in the latest PDG baryon tables are included in GENIE~\cite{Rein:2006di}, while NEUT considers all 18 resonances below the hadronic invariant mass of 2 GeV/c$^2$.
Axial mass values, $M_A^{RES}$, of 1.21 GeV/c$^2$ and 1.12 GeV/c$^2$ are used by NEUT and GENIE, respectively.

For GENIE, the Rein-Sehgal model is used up to a hadronic invariant mass of 1.7 GeV/c$^2$, while the analogous cutoff is 2 GeV/c$^2$ for NEUT.
Below this cut value, GENIE and NEUT use different descriptions of the nonresonant inelastic scattering. This is discussed further in Sec.~\ref{sec:BY}. 

\subsubsection{Coherent pion production }
Coherent scattering results in the production of forward going pions in both the 
 CC and NC channels.
Although both the NEUT and GENIE Monte Carlos use the Rein-Sehgal model~\cite{Rein:1982pf} for coherent pion production, they differ by a factor of 2 in total cross section, as GENIE uses a recent revision of the Rein-Sehgal model~\cite{Rein:2006di}. Both are, in turn, significantly larger than recent theoretical calculations~\cite{Dytman:2011zza}. The CC coherent pion production of pions has been observed for neutrino energies ranging from 2 to 80 GeV~\cite{Paschos:2005km}. At energies below 2 GeV, observations from K2K and SciBooNE are consistent with no coherent production~\cite{Hasegawa:2005td,Hiraide:2008eu}.

\subsubsection{Nonresonant inelastic scattering}\label{sec:BY}
There are similarities in how NEUT and GENIE treat nonresonant inelastic scattering. Both generators use the modifications suggested by Bodek \textit{et al}~\cite{Bodek:2004pc} to describe scattering at low $Q^2$.
In this model, higher twist and target mass corrections are accounted for through the use of a new scaling variable and modifications to the low $Q^2$ parton distributions.
The cross sections are computed at a fully partonic level and all relevant sea and valence quarks are considered. The longitudinal structure function is taken
into account using the Whitlow 
parametrization~\cite{Whitlow:1990gk}.
The default parameter values are those given in~\cite{Bodek:2004pc}, which are determined based on the GRV98 (Gl\"{u}ck-Reya-Vogt-1998)~\cite{Gluck:1998xa} parton distributions.
The same model can be extended to low energies, and it is used for the nonresonant processes that compete with resonances in the few-GeV region.

NEUT and GENIE use different methods to treat the nonresonant processes at low $W$ (i.e., low hadronic system invariant mass). Below 2 GeV/c$^2$, NEUT uses a $W$-dependent function to determine the pion multiplicity in each interaction.
The mean multiplicity of charged
pions is estimated from data collected in the Fermilab 15-foot bubble chamber experiment~\cite{Derrick:1977zi}. In both generators, Koba-Nielsen-Olesen (KNO) scaling~\cite{Koba:1972ng} is used to determine the charged hadron multiplicity.

Below 1.7 GeV/c$^2$, GENIE uses the Andreopoulos-Gallagher-Kehayias-Yang (AGKY) hadronization model~\cite{Yang:2009zx} to decompose the Bodek and Yang model into single pion and two pion production contributions.
A fraction of these CC-1$\pi$ and CC-2$\pi$ contributions are added to the Rein-Sehgal resonance model. The fractions are derived from fits to CC-inclusive, CC-1$\pi$ and CC-2$\pi$ bubble chamber data. The corresponding fractions for NC are worked out from the CC fractions using isospin arguments.

\subsubsection{Final state interactions }\label{sec:fsi}
Final state interactions (FSI) are strong interactions affecting particles produced in neutrino interactions as those particles traverse the target nucleus. These processes can produce a different observable state as compared to the state at the initial neutrino-nucleon interaction vertex. 

NEUT and GENIE use different microscopic cascade models to propagate the pion through the nuclear medium; the other hadrons including nucleons are also simulated with similar but simplified density-independent cascade models. 
In the case of pions, four processes are simulated as FSI: QE scattering, absorption, charge exchange and particle production; in the case of nucleons, the simulated processes are QE scattering as well as single and double pion production.

Because of the larger effect pion tracks have on the CC event selection, we describe pion interactions in more detail. In NEUT, the calculation of the probability of interaction is separated into low and high pion momentum regions ($\leq 500$ MeV/c and $>500$ MeV/c) where different models are adopted.
In the region $p_{\pi}<400$ MeV/c, the $\Delta$-hole model~\cite{Salcedo:1987md} is employed. For $p_{\pi} > 500$ MeV/c, $\pi p$ scattering cross sections are used to calculate the interaction probabilities.
For $p_{\pi}>400$ MeV/c and $p_{\pi} < 500$ MeV/c, a linear blending is done to alleviate discontinuities at 500 MeV/c between the $\Delta$-hole model and the scattering data.
GENIE makes use of the INTRANUKE model~\cite{Dytman:2009zz}, while NEUT uses the model described in~\cite{Hayato:2009}.
The FSI models in both generators are tuned to external pion-nucleon scattering data~\cite{dePerio:2011zz,Andreopoulos:2009rq}. 

\subsection{Neutrino interaction model uncertainties}\label{sec:modelsyst}
In this section, we explain how the cross-section uncertainties have been calculated.
We use a data-driven method
where the NEUT predictions are compared to 
available neutrino-nucleus data in the energy region relevant for
T2K. We fit the free parameters of the models implemented in
NEUT, and introduce \textit{ad hoc} parameters, often with large uncertainties,
to take account of remaining discrepancies between NEUT and the data.

\subsubsection{Estimation of the CCQE scattering uncertainty}
The uncertainty in the quasielastic cross section is estimated by comparing data from the MiniBooNE experiment to NEUT. In this comparison, 
NEUT CCQE interactions are simulated using the predicted MiniBooNE flux~\cite{AguilarArevalo:2008yp} and tuned to the MiniBooNE double-differential muon CCQE data~\cite{AguilarArevalo:2010zc} to fit for the best value of $M_A^{QE}$ and CCQE normalization. The error on each NEUT parameter is determined as the difference between the fitted value of the parameter and the nominal. 
For CCQE, we take the MiniBooNE data below 1.5 GeV, and assign the uncertainty of 11\% as reported by the MiniBooNE Collaboration~\cite{AguilarArevalo:2010zc}.  
To allow for the discrepancy in the CCQE cross section between NOMAD~\cite{Lyubushkin:2008pe} at $\sim$10 GeV and MiniBooNE at $\sim$1 GeV a 30\% error has been set above 1.5 GeV.

To explore the dependence of the result on the use of the simple RFG nuclear model, a spectral function (SF) model~\cite{Benhar:2005dj} implemented in the NuWro generator~\cite{Golan:2012wx} is used for comparison. The major contribution to the SF comes from the shell model and the remaining $\sim$20\% from correlated pairs of nucleons. The latter part accounts for a high momentum tail in the nucleon momentum distribution, which extends far beyond the Fermi momentum. The effective binding energy (equivalent to $E_B$ in the Fermi gas model) is on average larger which makes the cross section smaller for the SF implementation as compared to RFG. 
The fractional difference in CCQE event yields calculated with the RFG model and the SF is used to set the uncertainty in each bin of reconstructed muon momentum and angle.
Because the uncertainty on the Fermi momentum is large enough to have a non-negligible effect on the shape of the $Q^2$ spectrum for CCQE events, a systematic error on the Fermi momentum is added over and above the systematic uncertainty assigned using the spectral function comparison.
In this case, the uncertainty on the Fermi momentum is taken from electron scattering data~\cite{Smith:1972xh} and increased to cover MiniBooNE data at low $Q^2$. 

Table~\ref{tab:priorxsec_ccqe} summarizes the uncertainties related to the charged current quasielastic interactions for the NEUT generator.

\begin{table}[htb]
 
\caption{\label{tab:priorxsec_ccqe} Summary of the CCQE cross-section uncertainties used in this analysis. $x_1^{QE}$, $x_2^{QE}$ and $x_3^{QE}$ denote the CCQE normalization for different energy ranges, while $x_{SF}$ describes the nuclear model applied, where $x_{SF}=0$ corresponds to the use of the relativistic Fermi gas model and $x_{SF}=1$ to the use of the spectral function. }

\centering
\begin{tabular}{c c c c }
\hline
\hline
 Parameter & Energy range (GeV)  & Nominal value & Error \\
\hline
$M_{A}^{QE}$ &All $E_{\nu}$ & 1.21 GeV & 37\% \\
$x_1^{QE}$ &$0.0<E_{\nu} <1.5$  & 1& 11\% \\
$x_2^{QE}$ &$1.5<E_{\nu} <3.5$  & 1&30\% \\
$x_3^{QE}$ &$3.5<E_{\nu}$   & 1&30\% \\
$x_{SF}$ &All $E_{\nu}$ & 0 & 100\% \\
$p_F$ &All $E_{\nu}$ &  217 MeV/c & 14\% \\ 
\hline
\hline
\end{tabular}
\end{table}

\subsubsection{Estimation of the CC non-QE scattering uncertainty}
To constrain the single pion production, we perform a joint fit to the MiniBooNE data sets for CC-1$\pi^0$~\cite{AguilarArevalo:2010xt}, CC-1$\pi^+$~\cite{AguilarArevalo:2010bm} and NC-1$\pi^0$ production~\cite{AguilarArevalo:2009ww}, since these sets are connected in the underlying (Rein-Sehgal) model. We fit to the reconstructed $Q^2$ distributions in the CC channels and the pion momentum distribution in the NC channel.
Nine parameters, described in subsequent paragraphs, are included in the fit: $M_A^{RES}$, $W_{\rm{eff}}$, the CC coherent normalization ($x^{CCcoh.}$), CC-1$\pi$ normalizations ($x_1^{CC1\pi}$ and $x_2^{CC1\pi}$), CC-other shape ($x_{CCother}$), NC-1$\pi^{0}$ normalization ($x^{NC1\pi^0}$), NC-1$\pi^{\pm}$ normalization ($x^{NC1\pi^{\pm}}$), NC coherent normalization  ($x^{NCcoh.}$), and NC-other normalization ($x^{NCother}$). 
The MiniBooNE data cannot constrain all the parameters separately.
The fit can, however, reduce the total parameter space through correlations. Because the MiniBooNE data sets used in these fits cannot constrain the parameters $x_{CCother}$, $x^{NC1\pi^{\pm}}$,  $x^{NCcoh.}$ and $x^{NCother}$, we add penalty terms to prevent large excursions away from the generator defaults in the best fit values of these parameters. These terms represent conservative prior uncertainty estimates on these parameters and are derived with the use of K2K and SciBooNE measurements. 
 
The MiniBooNE data directly constrain $M_A^{RES}$, $x^{NC1\pi^{0}}$ and the CC-1$\pi$ normalization parameter for neutrino energies below 2.5 GeV, $x_1^{CC1\pi}$.
For energies above 2.5 GeV, we assign a conservative 40\% error to the normalization of CC-1$\pi$ production, motivated primarily by NOMAD data~\cite{Lyubushkin:2008pe}.
The $W_{\rm{eff}}$ parameter modifies the width of the hadronic resonance, but not its normalization. It allows an adjustment of the shape of the $|\vec{p}_{\pi^0}|$ spectrum of the NC-1$\pi^0$ channel to improve agreement with data. The error on this parameter is taken to be 50\%.

 A 100\% error has been set for the CC coherent pion production, $x^{CCcoh}$. This is driven by the fact that K2K and SciBooNE~\cite{Hasegawa:2005td,Hiraide:2008eu} data indicate there is much less coherent charged pion production by neutrinos with energy below 2 GeV than predicted by the original models.

The $x_{CCother}$ parameter modifies a combination of other CC cross-section channels as a function of $E_{\nu}$. The interactions contributing to this category are the CC-n$\pi$ production, which are interactions with more than one pion in the final state that have a hadronic mass between 1.3 GeV and 2 GeV, and DIS or CC resonant interactions with $\eta/K/\gamma$ production. 
From external data sets~\cite{Lyubushkin:2008pe}, the uncertainty is known to be of the order of 10\% at 4 GeV. Using this as a reference point, the error is defined as decreasing with the neutrino energy (0.4 GeV/$E_{\nu}$ ).

The NC-1$\pi^{\pm}$ and the NC-other normalization uncertainties are set to 30\% following the studies done for the first published T2K oscillation analyses~\cite{Abe:2011sj}. In MiniBooNE, there are very few events corresponding to these channels and the normalization of these events is not well constrained in the fit. 

A 30\% uncertainty on the NC coherent normalization factor is used.
This conservative estimate is motivated by the observation of a 15\% discrepancy between the NEUT prediction and the SciBooNE measurement of NC coherent production~\cite{Kurimoto:2010rc}, together with a 20\% systematic error in those data.

In addition to the nine parameters mentioned earlier, two additional parameters are considered: the $x_{1\pi E_{\nu}}$ parameter and the rate of $\pi$-less $\Delta$ decay, $x_{\pi-less}$. The $x_{1\pi E_{\nu}}$ parameter is an empirical parameter that covers the discrepancy between the MiniBooNE measurement of the CC-1$\pi^+$ cross section versus $E_{\nu}$ and the NEUT prediction using the best fit parameters from the fit to MiniBooNE data described above. The discrepancy is as large as 50\% at $E_{\nu} = 600$ MeV. 

$\pi$-less $\Delta$ decay, also called $\Delta$ reabsorption~\cite{Oset:1987re}, 
is the interaction of the $\Delta$ within the target nucleus prior to decay, yielding no pion. This process is assumed to occur in $\sim$20\% of resonant interactions~\cite{Singh:1998ha}. It is currently implemented in NEUT independently of energy and target, and results in a CCQE-like event. An absolute error of 20\% is assigned to this process.

Table~\ref{tab:priorxsec_res} summarizes the different uncertainties on the nonquasielastic channels for the NEUT MC.

\begin{table}[htpb]
\caption{\label{tab:priorxsec_res}
Summary of the CC non-QE cross-section uncertainties.}
\centering
\begin{tabular}{c c c c  }
\hline
\hline
 Parameter  & Energy range  & Nominal & Relative \\
  & (GeV) &  value & Error \\
\hline
$M_{A}^{RES}$ &All $E_{\nu}$ & 1.16 GeV & 9\% \\ 
$W_{\rm{eff}}$ &All $E_{\nu}$ & 1 & 52\% \\
$x^{CCcoh.}$ &All $E_{\nu}$ & 1& 100\% \\
$x_1^{CC1\pi}$ &$0.0<E_{\nu} <2.5$  & 1.63 & 26\% \\
$x_2^{CC1\pi}$ &$2.5<E_{\nu}$  & 1&40\% \\
$x_{CCother}$ &All $E_{\nu}$  & 0 & 40\% at 1 GeV \\
$x^{NC1\pi^0}$ &All $E_{\nu}$  &1.19& 36\% \\
$x^{NC1\pi^{\pm}}$  &All $E_{\nu}$  &1& 30\% \\
$x^{NCcoh.}$ &All $E_{\nu}$  &1& 30\% \\
$x^{NCother}$ &All $E_{\nu}$  &1& 30\% \\
$x_{1\pi E_{\nu}}$ &All $E_{\nu}$ & off & 50\% \\
$x_{\pi-less}$ &All $E_{\nu}$ & 0.2 & 100\% \\
\hline
\hline
\end{tabular}
\end{table}

\subsubsection{Estimation of the FSI uncertainty}\label{sec:fsierr}
In theory, the uncertainties on the FSI parameters (absorption, charge exchange, QE scattering and inelastic scattering) are correlated with the other cross-section parameters.
In this analysis, however, we assume them to be independent, as a first approximation. Therefore, the uncertainty on the FSI contribution is added in quadrature to the other sources in the reported cross section (see Sec.~\ref{sec:method}).
The uncertainties on the parameters that scale the microscopic interaction probabilities are shown in Table~\ref{tab:fsitab}. They have been estimated from comparison to external $\pi$-$^{12}{\mbox{C}}$ scattering data~\cite{dePerio:2011zz}.

\begin{table}[!h]
\caption{\label{tab:fsitab}
 A list of parameters used to calculate the FSI uncertainties for the NEUT generator only. The low and high momentum range refer to the pion momentum smaller and greater than 500 MeV/c (see Sec.~\ref{sec:fsi}).}
\centering
\begin{tabular}{c c c c  }
\hline
\hline
 Parameter   & Error \\
\hline
Absorption (low momentum) & 45\% \\
Charge exchange (low momentum) & 60\% \\
QE scattering  (low momentum)  & 60\% \\
Charge exchange (high momentum) & 30\% \\
QE scattering (high momentum)  & 40\% \\
Inelastic scattering (high momentum)  & 50\% \\
\hline
\hline
\end{tabular}
\end{table}

\section{Reconstruction and selection of charged current neutrino interactions}\label{sec:ccsel}

At the core of this analysis is the reconstruction and event selection in the MC and the data, as described in detail in this section. In this analysis, we select CC-inclusive candidates by identifying negatively charged muon-candidate tracks in the tracker. The number of selected events in reconstructed bins is used to infer the number of true events and, consequently, the CC-inclusive cross section, as described in Sec.~\ref{sec:ccxs}. The selected events in each bin are dependent on the reconstruction and detection efficiency. 
Below, in Sec.~\ref{sec:detsim}, we first review how the various pieces of MC simulation are included in the final MC. Sec.~\ref{sec:reco} explains in more detail how the tracks are reconstructed in the TPCs and FGDs. The event selection, described in Sec.~\ref{sec:sel}, relies on these reconstructed tracks. The performance of the event selection and reconstruction is provided in Sec.~\ref{sec:perf} along with a data-MC comparison. Finally, a discussion of the detector response systematic uncertainty is given in Sec.~\ref{sec:detsys}, while Sec.~\ref{sec:ccxs} describes how the detector uncertainties are then propagated to the final cross section.

\subsection{Monte Carlo simulation}\label{sec:detsim}
The MC simulation of neutrino interactions in the near detector can be divided into several steps. First, the neutrino flux is simulated and propagated to the near detector (see Sec.~\ref{sec:flux}). The neutrino interactions in ND280 are then simulated by the two generators NEUT and GENIE (see Sec.~\ref{sec:nuint}), while only NEUT is used to simulate interactions located outside ND280, e.g. the pit wall, sand and magnet. The secondary interactions in ND280, the response of the active detector components and readout electronics are simulated using a mix of GEANT4~\cite{Agostinelli:2002hh} and custom software.
 The MC statistics used for this analysis correspond to 1.7$\times 10^{21}$ POT for interactions in ND280 (for both NEUT and GENIE) and 7.0$\times 10^{19}$ POT for interactions outside ND280.

\subsection{Track reconstruction}\label{sec:reco}
The TPC detector was previously discussed in Sec.~\ref{sec:nd280oa}.
The first step in the TPC reconstruction is the application of the gain calibration constants and the removal of dead and noisy pads, resulting in a waveform representing the charge acquired in a single pad as a function of the readout time. The charge and time of pulses observed in the waveform are extracted, and clusters are formed by joining coincident pulses on adjacent pads in vertical columns. Contiguous clusters are then joined to form track candidates and the track parameters are obtained with a maximum-likelihood fit to the observed charge distribution assuming a helical trajectory and accounting for the diffusion of the electrons ionized by the track. The drift distance is calculated by extracting the time for the passage of the track ($t_0$) by matching the track to objects reconstructed in the fast scintillator-based detectors, i.e., the FGDs, P\O{}D and ECals.

To perform the particle identification (PID) on a track, the charge in the clusters is first corrected for the distance traveled by the ionization electrons. This distance is estimated based on the reconstructed track position. A truncated mean charge is then formed from the corrected cluster charges. The expected charge deposition for each of several particle hypotheses is calculated for the measured momentum of the track, and compared to the measured charge.

The FGD data is reconstructed after the TPC reconstruction.
First the FGD hits are separated into various clusters in time.
 The TPC tracks are then matched to FGD hits to form a three-dimensional track.

\subsection{Event selection}\label{sec:sel}
The event selection consists of a series of cuts designed to
select $\nu_{\mu}$ CC-inclusive interactions in FGD1.
Interactions in FGD2 are not considered in this paper, as they involve a more complex target consisting of a mix of carbon and oxygen.
An extension of this analysis to FGD2 interactions is anticipated in the future.
In addition, there is no attempt in this analysis
to select backward-going muons. Therefore, the CC-inclusive selection cuts that follow are based on the observation of forward-going tracks compatible with negatively charged muons. The cuts used in this analysis are described below.
\begin{enumerate}
\item \textit{Data quality flag}\vspace{0.15cm}\\
 We require that the whole ND280 off-axis detector in a full spill is working properly. 
\item \textit{Time bunching}\vspace{0.15cm}\\
Tracks are grouped together in bunches according to their times. This treats neutrino interactions in two different bunches within the same beam spill as two different events, reducing the accidental pileup of events. 
The bunch width is $\sim$ 15.0 ns in data.
 Tracks are associated within a bunch if they deviate from the mean bunch time by less than 60 ns (i.e. four times the bunch width in data), other tracks are removed. 

\item \textit{Negatively charged track in the FGD1 fiducial volume}\vspace{0.1cm}\\
We require at least one negatively charged track (with FGD and TPC components) starting inside FGD1's fiducial volume with more than 18 TPC clusters.
 The interaction vertex is defined as the beginning of this track described by the coordinates $(x_0,y_0,z_0)$. It corresponds, in general, to the place where the 3d-fitted track intercepts the vertical plane of the upstream-most matched FGD hit.

The FGD1 volume and fiducial volume are shown in Fig.~\ref{fig:fgd}.
 As mentioned earlier, five bars on either end of each layer in the FGD are excluded from the fiducial volume in the $x$ and $y$ dimensions, while the upstream $z$ cut places the fiducial volume just after the first $xy$ module. The fiducial volume contains therefore 14 $xy$ modules in which the $x$ and $y$ layers contain 182 scintillator bars. 
With this definition, the fiducial volume cut becomes $|x_0|<874.5$ mm, $|y_0-55~\rm{mm}|<874.5$ mm, and $z_0 \in [136.9,447.0]$ mm.

 The requirement that the track should contain at least 18 clusters is called the \textit{TPC track quality cut} and it rejects short tracks for which the momentum reconstruction and particle identification is less reliable. The choice of this particular value of the quality cut is based on studies of the kinematic bias for tracks of different length. Since only a small fraction of the selected tracks have fewer than 19 hits, the effect of the quality cut, and the systematic error associated with it, is very small.

If there is more than one negatively charged track passing these cuts, we select the highest momentum track as the muon candidate.

\item \textit{Upstream veto}\vspace{0.1cm}\\
The goal of this cut is to remove events entering the FGD1 fiducial volume from the upstream face of the detector. Events with more than one reconstructed track are rejected if there is a track starting more than 150 mm upstream of the muon candidate starting position.

\item \textit{TPC particle identification (PID)}\vspace{0.1cm}\\
 Given the estimated momentum of the muon candidate, a discriminator function is calculated for the muon, pion, and proton hypotheses based on the energy loss of the track in the TPC. This cut, applied to the muon candidate, rejects electrons at low momentum (below 500 MeV/c) and removes protons and pions. 
\end{enumerate}

\begin{figure}[!h]
    \centering
    \includegraphics[width=0.45\textwidth]{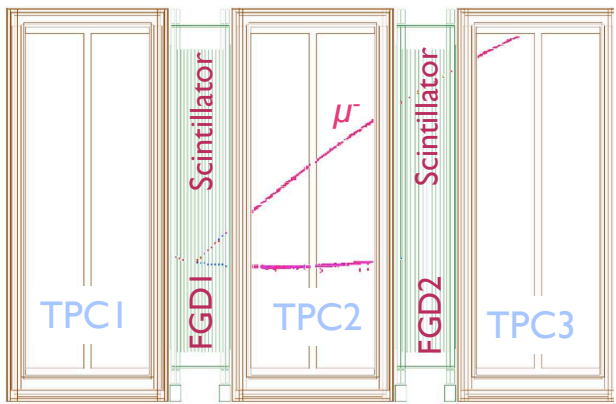}
    \caption{Side view of a charged current event candidate in the tracker region of the near detector, which shows the bending due to the magnetic field. The muon candidate is reconstructed with an angle of 40$^\circ$ and a momentum of 566 MeV/c.}
    \label{fig:ccevent}
  \end{figure}

Taken together, the cuts above define the CC-inclusive selection in FGD1. The events surviving these cuts are included in the final data sample for further analysis. Figure~\ref{fig:ccevent} shows one of the events selected via these cuts.

\subsection{Performance}\label{sec:perf}

Table~\ref{tab:sel} shows the number of selected events in the data and the prediction of the two generators, after each cut.
\begin{table}[!h]
\caption{\label{tab:sel}
The number of selected events for the different cuts as predicted by the NEUT and GENIE generators compared to data. The numbers are normalized to the data POT of 10.8$\times 10^{19}$.
}
\centering
\begin{tabular}{c | c   c    c  }
\hline
\hline
Cut & Data & NEUT & GENIE \\
  \hline
Good negative track in FV & 8837 & 8899 & 8673 \\
Upstream veto & 6243 & 6582 & 6351\\
PID cut & 4485 & 4724 & 4536\\
\hline
\hline
\end{tabular}
\end{table}

In Table~\ref{tab:effpur}, the resulting inclusive charged current efficiency and purity are shown for each cut. The primary effect of the upstream veto cut is to remove events taking place in the P\O{}D or in magnet coils that cause a secondary interaction in the FGD. 
The increase of purity after the PID cut is due to the PID discriminator's ability to separate muons from the low momentum electrons produced outside the fiducial volume by mostly NC events. 
As a consequence of the selection, the muon candidate in the signal sample is identified correctly 96\% of the time.

\begin{table}[!h]
\caption{\label{tab:effpur}
The CC efficiency, $\epsilon$, and purity, $\mathcal{P}$, estimated with the MC simulation.
}
\centering
\begin{tabular}{c | c c | c c }
\hline
\hline
\multirow{2}{*}{ Cut} & \multicolumn{2}{c|}{NEUT} &\multicolumn{2}{c}{GENIE} \\
 &  $\epsilon$ (\%) & $\mathcal{P}$ (\%) &  $\epsilon$ (\%) & $\mathcal{P}$ (\%) \\
\hline
Good negative track in FV &  56.8 & 52.7    & 58.3 & 51.1
\\
Upstream veto &  54.4 & 68.4 & 56.2 & 67.4
\\
PID cut &  49.5 & 86.8 & 51.2 & 86.2
\\
\hline
\hline
\end{tabular}
\end{table}

\begin{figure}[!h]
    \centering
    \includegraphics[width=0.5\textwidth]{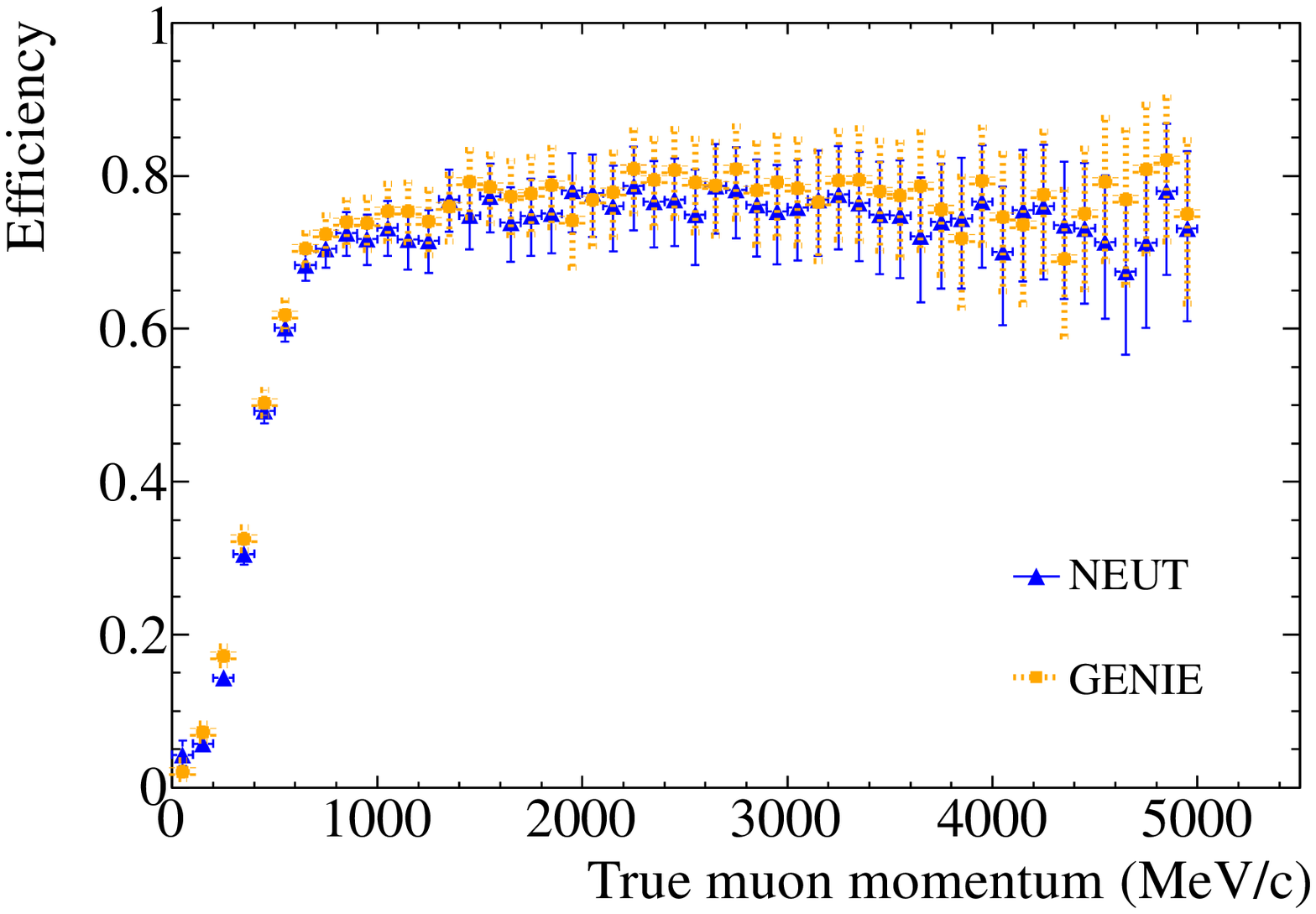}
    \caption{The event selection efficiency as a function of the muon momentum with its statistical error bars.}
    \label{fig:eff0}
    \includegraphics[width=0.5\textwidth]{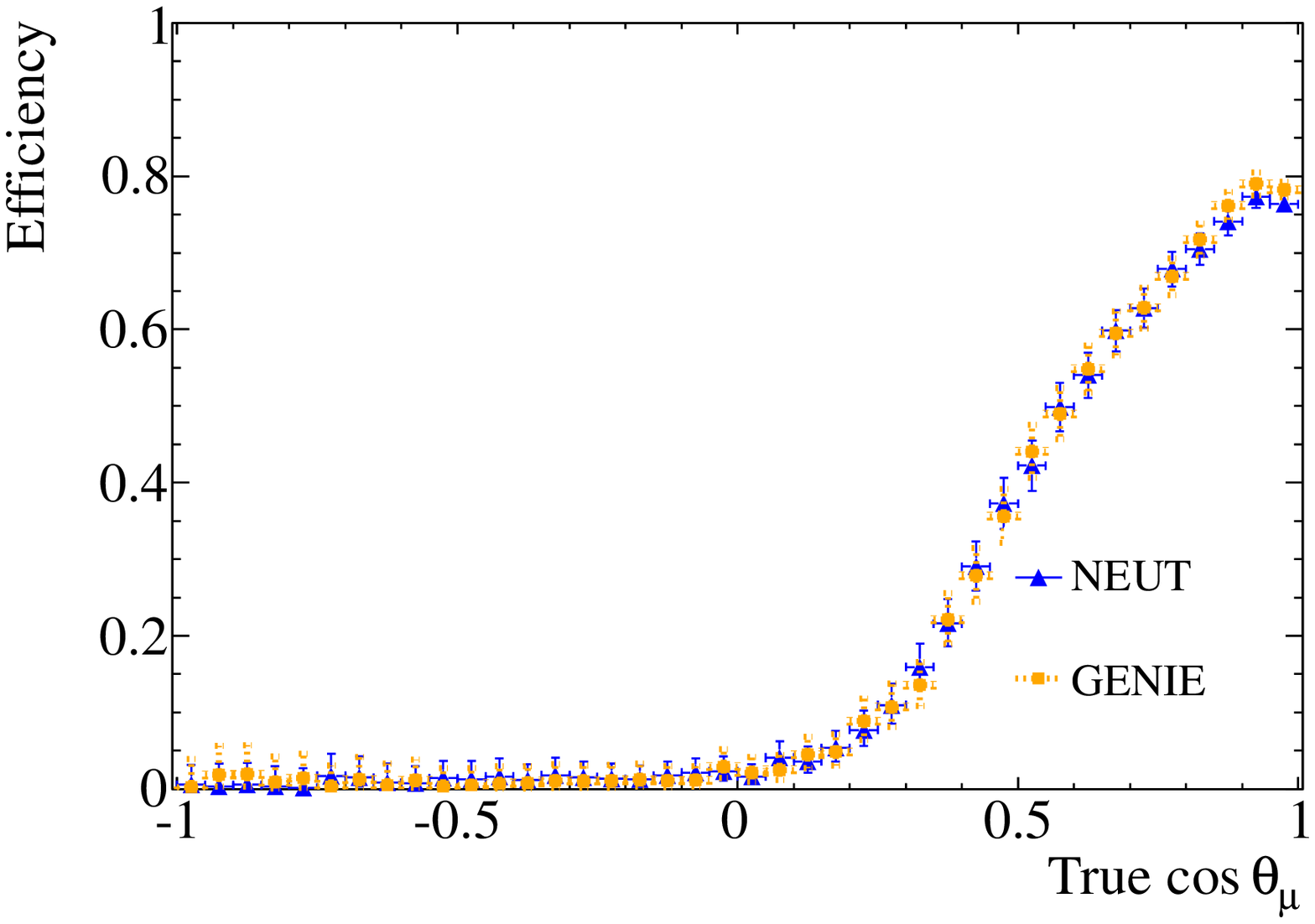}
    \caption{The event selection efficiency as a function of the muon angle with its statistical error bars. The nonzero efficiency for the backward-going muon is due to misreconstruction.}
    \label{fig:eff}
  \end{figure}

In Figs.~\ref{fig:eff0}~and~\ref{fig:eff}, the efficiency as a function of muon momentum and angle are shown as estimated from NEUT and GENIE MC simulations.
The efficiency for backward-going muons is very low since the reconstruction does not attempt to reconstruct backward-going tracks. 
The nonzero efficiency for backward-going muons arises from the reconstruction of events with a forward-going pion possessing a negative charge that is misidentified as a negatively charged muon, while the real muon is going backwards.
This effect is included in the cross-section determination of Sec.~\ref{sec:syst}.

The various backgrounds in the selected sample are shown in Table~\ref{tab:bg}. The main background comes from interactions outside the FGD (external background). 

\begin{table}[!h]
\caption{\label{tab:bg}
 The background composition of the CC-inclusive selection according to the NEUT and GENIE MC generators.  }
\centering
\begin{tabular}{c | c  c }
\hline
\hline
\multirow{2}{*}{ Type}  &\multicolumn{2}{c}{Fraction of sample (\%)} \\
& NEUT & GENIE \\
\hline
Outside FV but in FGD1 & 0.94 $\pm$ 0.14 & 0.91 $\pm$ 0.14 \\
Outside FGD1 & 8.16  $\pm$ 0.40 & 8.36 $\pm$ 0.40 \\
Neutral Currents in FV & 3.17 $\pm$ 0.26 & 3.59 $\pm$ 0.28\\
 CC$\nu_e$ in FV & 0.27 $\pm$ 0.08 & 0.26 $\pm$ 0.08    \\
$\bar{\nu}_{\mu,(e)}$ in FV & 0.68 $\pm$ 0.12 & 0.71 $\pm$ 0.12\\
\hline
Total & 13.2 $\pm$ 0.5 & 13.8 $\pm$ 0.5\\
\hline
\hline
\end{tabular}
\end{table}

\begin{figure}[hpbt]
\begin{center}
  \includegraphics[width=0.49\textwidth]{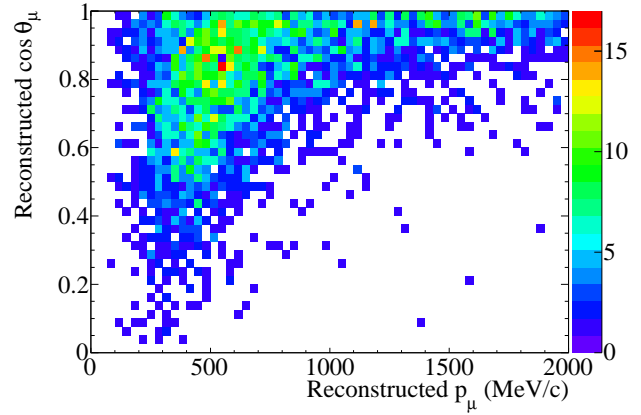} 
  \caption{The distribution of selected events in the data in the ($p_{\mu},\cos\theta_{\mu}$) plane.}
\label{fig:pth}
\end{center}
\end{figure}

\begin{figure}[hptb]
\begin{center}
  \includegraphics[width=0.49\textwidth]{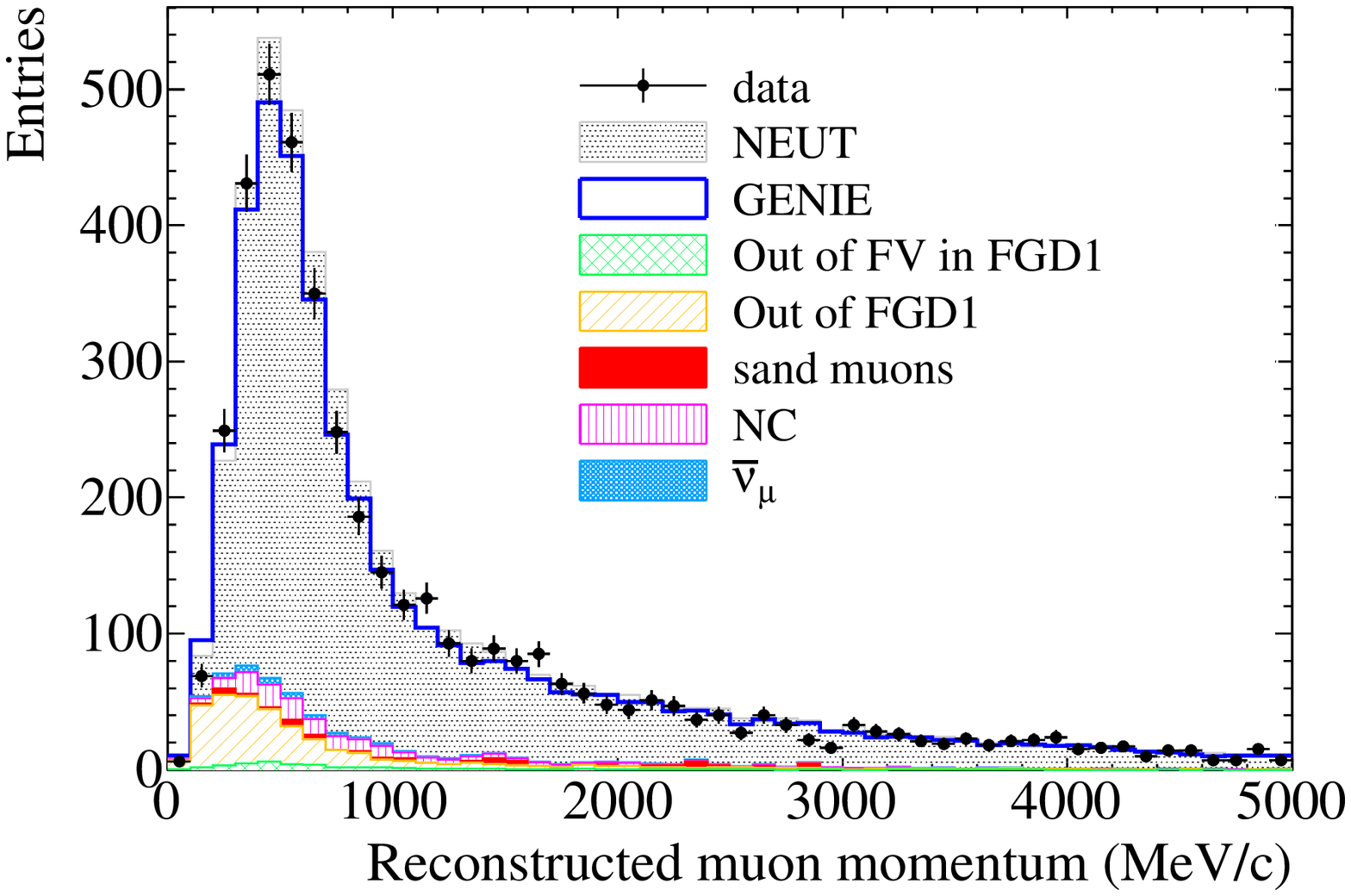} 
  \includegraphics[width=0.49\textwidth]{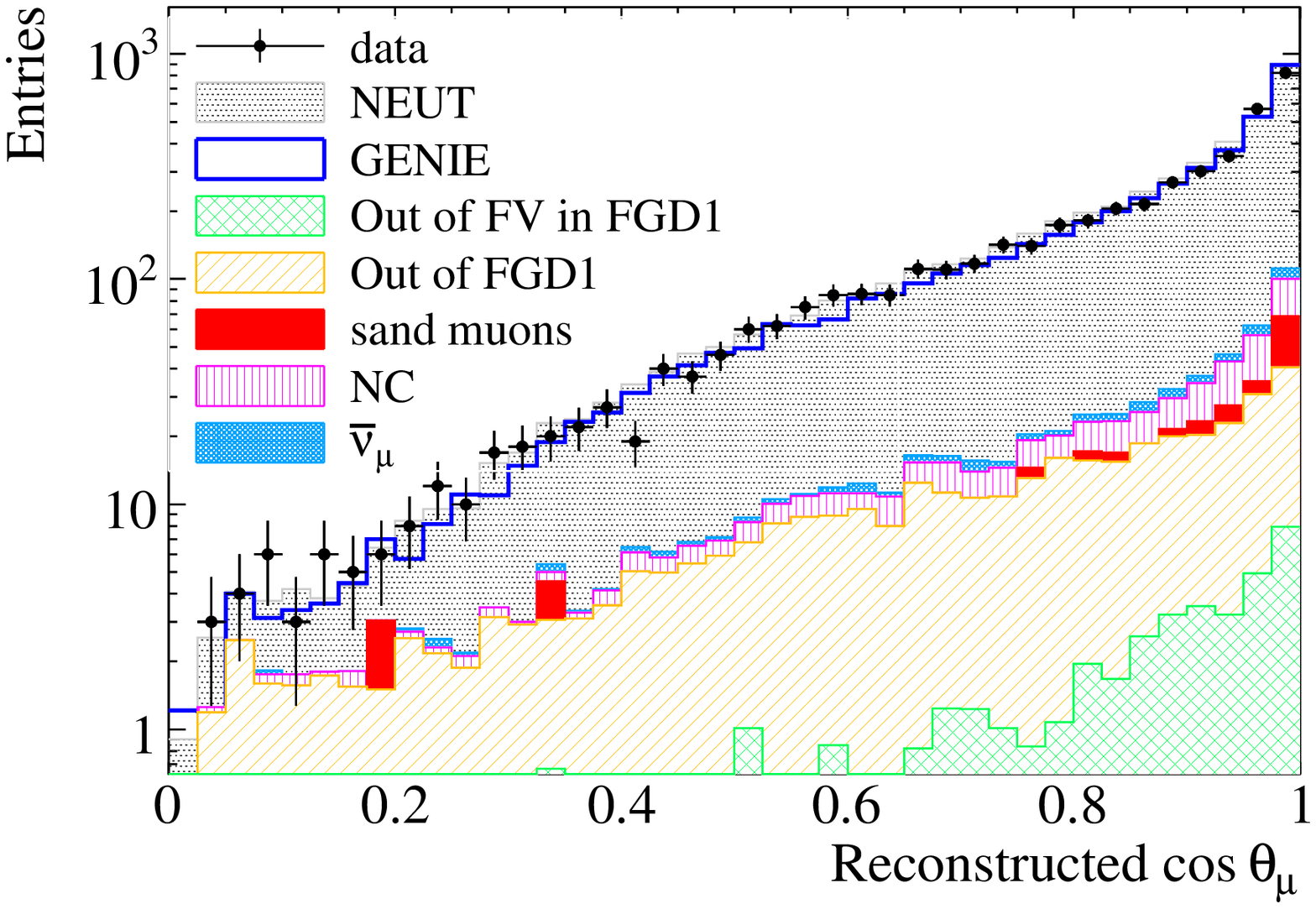} 
  \caption{The muon momentum and angle distribution for the selected events in the data and MC. The GENIE prediction is shown by the solid blue line, while the backgrounds and signal as derived from NEUT  are shown in filled colors. }
\label{fig:thmom}
\end{center}
\end{figure}

The phase space for events selected in the data is shown in Fig.~\ref{fig:pth}. The momentum and angular distributions are shown in Fig.~\ref{fig:thmom}, where the level of the various backgrounds is taken from the NEUT Monte Carlo and the overall GENIE prediction is superimposed.

\subsection{Detector response uncertainties}\label{sec:detsys}
The systematic errors on the detector response can be separated into two categories: the uncertainties on the backgrounds in the selected sample and the uncertainties in the reconstruction. Depending on the uncertainty source, the systematic error is computed by data/MC comparison, the use of special Monte Carlo calculations, the study of cosmic ray muons, or external measurements. 
For each source, a covariance matrix on the predicted number of events is computed according to the two-dimensional binning in momentum and angle described in Sec.~\ref{sec:method}. 
The two main uncertainties due to the detector response come from the momentum uncertainty arising from distortions in the magnetic field and from external backgrounds.

The magnetic field was mapped for different magnet currents in 2009 and 2010~\cite{Abe:2011ks}. In addition, a pattern of thin aluminum discs and strips were glued to the cathode in the TPC. By the use of a laser emitting 266 nm light, electrons from the aluminum are ejected and measured in the TPC~\cite{T2KtpcNIM2010}. The displacement of the electrons is used to estimate the magnetic field distortions. Depending on the bin, the size of this uncertainty varies from 0.3\% to 7\% on the predicted number of events.

The external background uncertainty is evaluated by dividing it into two main categories: the external background coming from outside FGD1 and the background coming from inside FGD1 but outside the fiducial volume. 
A 20\% uncertainty is assigned based on a comparison of the interaction rate of the background outside FGD1 between data and Monte Carlo.
A reconstruction uncertainty is also assigned for several categories of events such as those containing tracks with very high angles, tracks in which two FGD hits are missing or tracks where the matching reconstruction has failed. The systematic uncertainty is assigned based on the difference in the failure rate between the data and MC for each category of reconstruction considered.
The total uncertainty assigned to the external background varies from 0.4\% to 9\%, depending on the bin.

A momentum scale uncertainty, determined using the aforementioned magnetic field measurements, is included in the analysis. The momentum resolution uncertainty is estimated by comparing the momentum difference in different TPCs in data and MC for tracks crossing at least two TPCs.

 Reconstruction uncertainties associated with TPC-FGD matching efficiency, tracking efficiency and hit efficiency are also evaluated. 
 The TPC-FGD matching efficiency is calculated for each bin separately using control samples such as cosmic ray tracks passing through TPC2 and FGD1. The error on this efficiency varies from 0.2\% to 2\%, leading to an uncertainty on the predicted number of events smaller than 1\% in all bins. 
The uncertainty on the tracking efficiency is determined by comparing tracks with high muon purity crossing the entire detector in data and MC. It is found to be 0.5\%, leading to an uncertainty smaller than 0.8\% on the predicted number of events. 
The hit efficiency uncertainty is estimated to be 0.1\% by comparing the distribution of the number of clusters for data and MC. This uncertainty leads to one of the smaller uncertainties on the predicted number of events (below 0.002\%).

An additional source of error arises when the charge of a track is not determined properly.
 A charge confusion probability is computed for data and MC using a sample of events with tracks starting in the P\O{}D and crossing at least two TPCs by looking at the fraction of events for which the charges differ in the different TPCs. The 0.3\% data/MC difference leads to less than 1.1\% uncertainty on the predicted number of events in each bin.

The uncertainty on the particle identification is obtained by using samples with a high purity of muons. Generally, these are tracks starting in the P\O{}D which cross the three TPCs. For such tracks, the difference of the energy loss from the expectation is computed for data and MC. The uncertainty is then based on the difference between data and MC leading to an uncertainty smaller than 0.6\% in all bins.

Finally, the uncertainties due to pileup events arising from neutrino interactions in the near detector and in the sand are estimated to be 0.2\% and 1.5\%, respectively. The uncertainty from cosmic-ray-induced pileup events is estimated to be 0.1\%. The effect of cosmic rays on the number of predicted events is negligible.

Table~\ref{tab:detsystprior} shows all the detector uncertainties studied for this analysis.

\begin{table}[!htb]
\caption{\label{tab:detsystprior}
A summary of all the systematic errors associated with the reconstruction and backgrounds. The first column lists all the uncertainties taken into account for the CC analysis, the second column presents the samples used to estimate the uncertainty and the last column shows the error size on the predicted events, depending on the bin. The uncertainties due the various backgrounds are shown at the bottom of the table.  }
\centering
\begin{tabular}{c c c  }
\hline
\hline
 Systematic Error & Data Sample  & Error (\%)  \\  
\hline
TPC momentum distortion& Special MC&  $0.3-7$ \\
TPC momentum scale& External data&  $0.1-2.4$ \\
TPC momentum resolution& Beam data/MC& $ 0.2-2.3$ \\
TPC-FGD matching efficiency& Sand muon + cosmics&  $ 0.2-1$ \\
TPC track efficiency&  Beam data/MC&   $0.05-0.8 $\\
Hit efficiency&  Beam data/MC&   $<0.002$ \\
Charge mis-ID& Beam data/MC&  $0.2-1.1 $ \\
TPC particle ID (PID)& Beam data/MC&  $0.02-0.6$  \\
\hline
External background& Several samples&  $0.4-9 $ \\
Sand muon background& Special MC&   $0.1-1.1 $ \\
ND280 pileup background & Beam data/MC &  0.2  \\
Cosmic ray background & Special MC&  Negligible  \\
\hline
\hline
\end{tabular}
\end{table}

\section{Flux-averaged charged current cross-section analysis}\label{sec:ccxs}
In what follows, the methodology for extracting the cross section from the selected events is described. Section~\ref{sec:method} summarizes how the central values are determined, while Sec.~\ref{sec:syst} provides the details of how the uncertainties were determined.

\subsection{Method} \label{sec:method}
To calculate the flux-averaged CC cross section, we use a method based on Bayes' theorem \cite{DAgostini} to unfold the number of reconstructed and selected events in each momentum and angle bin. The result of this unfolding gives the number of inferred events $\widehat{N}_k$ in ``true'' bins $k$, 
while the number of selected events in a reconstructed bin $j$ is $N_j^{\rm{sel}}$.
As demonstrated in \cite{DAgostini}, reconstructed and true bins do not necessarily have to be the same. 
As seen previously, this analysis does not attempt to reconstruct backward-going muons and all events in the $\cos\theta$ interval between $-1$ and 0.84 have been placed into a single bin (see Table~\ref{tab:PThmubinning}). 
As shown in Fig.~\ref{fig:eff}, there is almost no efficiency for the backward-going angle. We therefore split the true $\cos\theta$ bins into a truly backward bin [$-1$,0] and a more forward bin [0,0.84].
We will determine the double differential cross section in the forward direction only, while the total cross section is extrapolated into the backward direction using the MC and Bayes' theorem.
The binning and its association to the one-dimensional index are shown in Table~\ref{tab:truebinning}.

\begin{table}[h]
\caption{\label{tab:PThmubinning} Reconstructed muon momentum and angle binning with the correspondence to the one-dimensional binning index.}
\centering
\begin{tabular}{c|c c c c c }
\hline
\hline
$\cos\theta_{\mu}$ & \multicolumn{5}{c}{Reconstructed index number}     \\
\hline
$[0.94,1]$                  &   3  &    7   &   11    &   15  &19  \\
$[0.9,0.94]$                  &  2   &   6    &   10    &   14 & 18   \\
$[0.84,0.9]$                  &  1   &  5    &    9   &   13   & 17 \\
$[-1,0.84]$                  &   0  &   4    &   8    &    12  & 16 \\
\hline
$p_{\mu} (\mathrm{GeV}/c) $         &$[0,0.4]$&$[0.4,0.5]$&$[0.5,0.7]$&$[0.7,0.9]$& $[0.9,30]$ \\
\hline
\hline
\end{tabular}
\end{table}

\begin{table}[h]
\caption{\label{tab:truebinning} The true muon momentum and angle binning with the correspondence to the one-dimensional binning index.}
\centering
\begin{tabular}{c|c c c c c }
\hline
\hline
$\cos\theta_{\mu}$ & \multicolumn{5}{c}{True index number}     \\
\hline
$[0.94,1]$                  &   4  &    9   &   14    &   19  &24  \\
$[0.9,0.94]$                  &  3   &   8    &   13    &   18 & 23   \\
$[0.84,0.9]$                  &  2   &  7    &    12   &   17  & 22 \\
$[0,0.84]$                  &   1  &   6    &   11    &    16 & 21 \\
$[-1,0]$                  &   0  &   5    &   10    &  15  & 20 \\
\hline
$p_{\mu} (\mathrm{GeV}/c) $         &$[0,0.4]$&$[0.4,0.5]$&$[0.5,0.7]$&$[0.7,0.9]$& $[0.9,30]$ \\
\hline
\hline
\end{tabular}
\end{table}
\vspace{0.5cm}

The flux-averaged cross section in the one-dimensional bin, $k$, defined in Table~\ref{tab:truebinning}, is given by
\begin{eqnarray}
\langle \sigma_k \rangle_{\phi}=\frac{\widehat{N}_k}{T\phi}=\frac{1}{T\phi}\frac{\displaystyle \sum_j^{n_r} U_{jk}(N_j^{\rm{sel}}-B_j)}{\epsilon_k}\label{eq:xs}
\end{eqnarray}
 where $T$ is the number of target nucleons, $\phi$ is the integrated flux, $\widehat{N}_k$ is the number of inferred events, $U_{jk}$ is the unfolding matrix, $N_j^{\rm{sel}}$ is the number of selected events in the reconstructed bin $j$, $B_j$ the number of selected background events in this bin as predicted by the MC simulation. The parameter $n_r$ is the number of reconstructed bins and $\epsilon_k$ is the efficiency in the true bin $k$ predicted by the MC simulations. 

The unfolding matrix gives the probability that an event was created in bin $k$ given that it was reconstructed in bin $j$, and is defined according to Bayes' theorem as
\begin{eqnarray}
U_{jk}\equiv P(k|j)=\frac{P(j|k)P_0(k)}{ \displaystyle \sum_{\alpha} ^{n_t}P(j|{\alpha}) P_0({\alpha})}\label{eq:U}
\end{eqnarray}
where $n_t$ is the number of true bins and $P_0(k)$ is the probability to have a CC interaction in the true bin $k$, 
\begin{eqnarray}
P_0(k)=\frac{N_{k}}{N_{\rm{tot}}}\label{eq:P0}
\end{eqnarray}
where $N_{\rm{tot}}$ is the total number of events generated and $N_k$ the ones generated in the true bin $k$. $P_0(k)$ is given in Table~\ref{tab:P0} together with the efficiency in each bin.
Note that the model dependences entering the analysis are concentrated mainly in the background prediction and in the probability $P_0(k)$.

\begin{table}[tb]
 \caption{\label{tab:P0} The efficiency, $\epsilon_k$, and probability distribution for CC events, $P_0(k)$ as simulated by the nominal NEUT MC for each true momentum and angle bins.}
 \begin{tabular}{c c| c c  }
 \hline
 \hline
$P_{\mu}$ (GeV/c) & 	 $\cos\theta_{\mu}$  & $\epsilon_k$ (\%) & $P_0(k)$ (\%) \\
 \hline
  $[0.0,0.4]$ 
 	 & $[-1,0]$ 	   & 	 1.2 & 	 13.9 \\ 
 	 & $[0,0.84]$ 	   & 	 26.0 & 	 18.5 \\ 
 	 & $[0.84,0.90]$ 	   & 	 62.1 & 	 1.1 \\ 
 	 & $[0.90,0.94]$ 	   & 	 60.3 & 	 0.7 \\ 
 	 & $[0.94,1]$ 	   & 	 56.0 & 	 0.7 \\ 
 \hline 
   $[0.4,0.5]$ 
 	 & $[-1,0]$ 	   & 	 3.0 & 	 0.9 \\ 
 	 & $[0,0.84]$ 	   & 	 45.6 & 	 9.3 \\ 
 	 & $[0.84,0.90]$ 	   & 	 78.1 & 	 0.9 \\ 
 	 & $[0.90,0.94]$ 	   & 	 83.1 & 	 0.5 \\ 
 	 & $[0.94,1]$ 	   & 	 84.2 & 	 0.6 \\ 
 \hline 
   $[0.5,0.7]$ 
 	 & $[-1,0]$ 	   & 	 7.2 & 	 0.1 \\ 
 	 & $[0,0.84]$ 	   & 	 55.1 & 	 10.5 \\ 
 	 & $[0.84,0.90]$ 	   & 	 78.4 & 	 2.1 \\ 
 	 & $[0.90,0.94]$ 	   & 	 82.4 & 	 1.4 \\ 
 	 & $[0.94,1]$ 	   & 	 85.5 & 	 1.5 \\ 
 \hline 
   $[0.7,0.9]$ 
 	 & $[-1,0]$ 	   & 	 28.3 & 	 0.0 \\ 
 	 & $[0,0.84]$ 	   & 	 61.7 & 	 3.5 \\ 
 	 & $[0.84,0.90]$ 	   & 	 74.2 & 	 1.3 \\ 
 	 & $[0.90,0.94]$ 	   & 	 79.3 & 	 1.0 \\ 
 	 & $[0.94,1]$ 	   & 	 87.5 & 	 1.3 \\ 
 \hline 
   $[0.9,30.0]$ 
 	 & $[-1,0]$ 	   & 	 0.0 & 	 0.0 \\ 
 	 & $[0,0.84]$ 	   & 	 63.9 & 	 4.0 \\ 
 	 & $[0.84,0.90]$ 	   & 	 73.4 & 	 3.5 \\ 
 	 & $[0.90,0.94]$ 	   & 	 76.6 & 	 4.2 \\ 
 	 & $[0.94,1]$ 	   & 	 75.0 & 	 18.6 \\ 
 \hline 
 
\hline
\hline
\end{tabular}
\end{table}

$P(j|\alpha)$ and $P(j|k)$ are the probabilities to have an event reconstructed in the bin $j$ when it has been generated in the true bin $\alpha$ or $k$, respectively. $P(j|k)$ is estimated using the MC simulation as described in Sec.~\ref{sec:detsim} and is defined as the number of CC events reconstructed in bin $j$ and generated in bin $k$, $S_{jk}$, divided by the number of interactions generated in the true bin $k$, $N_k$, 
\begin{eqnarray}
P(j|k)=\frac{S_{jk}}{N_k}\label{eq:Prt}.
\end{eqnarray}
Note that $N_k$ contains all the CC events that were correctly selected as well as those that were missed by the selection.

\begin{figure}[!hptb]
\begin{center}
   \includegraphics[width=0.49\textwidth]{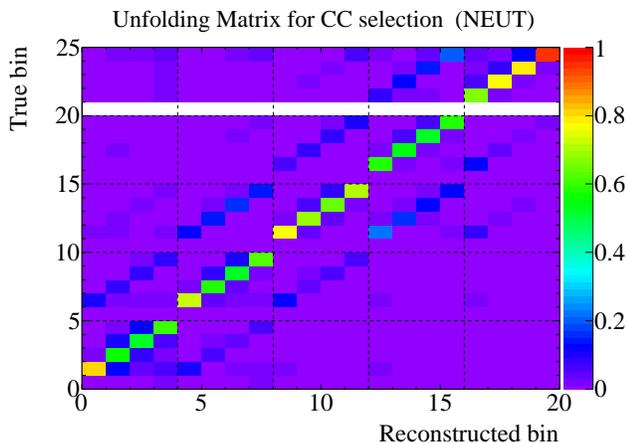}
  \caption{The unfolding matrix obtained with the NEUT MC simulation.
The white line corresponds to backward-going muons in the highest energy bin, for which our MC prediction is zero events. The $x$-axis and $y$-axis bin numbering is defined in Table~\ref{tab:PThmubinning} and Table~\ref{tab:truebinning} respectively.}
\label{fig:U}
\end{center}
\end{figure}

Applying the definitions in Eqs.~(\ref{eq:U}), (\ref{eq:P0}) and (\ref{eq:Prt}), the unfolding matrix can be rewritten as
 \begin{eqnarray}
U_{jk}=\frac{S_{jk}}{\sum_{\alpha}^{n_t}S_{j\alpha}}
\end{eqnarray}
and is shown in Fig.~\ref{fig:U}.

\subsection{Cross-section uncertainties}\label{sec:syst}
\subsubsection{Statistical error}
The statistical error is computed with events simulated using NEUT. The nominal NEUT MC is fluctuated $1000$ times, following Poisson statistics. The fluctuation is done simultaneously for $N^{\rm{sel}}_j$, $S_{jk}$, $B_j$ and the number of missed events needed for the efficiency. In this case, $N^{\rm{sel}}_j$ is the number of events selected as predicted by the MC, scaled down to the data POT and fluctuated accordingly. The error is then the rms of the fluctuated sample distribution, and contains the effect of the limited MC statistics. The size of the error varies from 4\% to 11\%, depending on the bin, and
the corresponding covariance matrix is given by
\begin{eqnarray}
V_{kl}=\frac{1}{1000}\sum_{s=0}^{1000}(\sigma_k^s-\sigma_k^{\rm{nom}})(\sigma_l^s-\sigma_l^{\rm{nom}}),
\end{eqnarray}
where $\sigma^s_k$ is the cross-section result obtained following Eq.~(\ref{eq:xs}) for the $s$th fluctuated sample, and $\sigma^{\rm{nom}}_k$ the result obtained with the nominal MC.
In addition, this method has been checked with analytic calculations.

\subsubsection{Systematic error}
The sources of systematic error on the cross section are the flux, neutrino interaction modeling, final state interactions, detector response, unfolding method and the knowledge of the number of target nucleons. The modeling of the final state interactions is treated here as independent of the rest of the neutrino interaction modeling, as discussed in Sec.~\ref{sec:fsierr}.

\begin{figure}[h]
  \centering
  \includegraphics[width=1 \columnwidth]{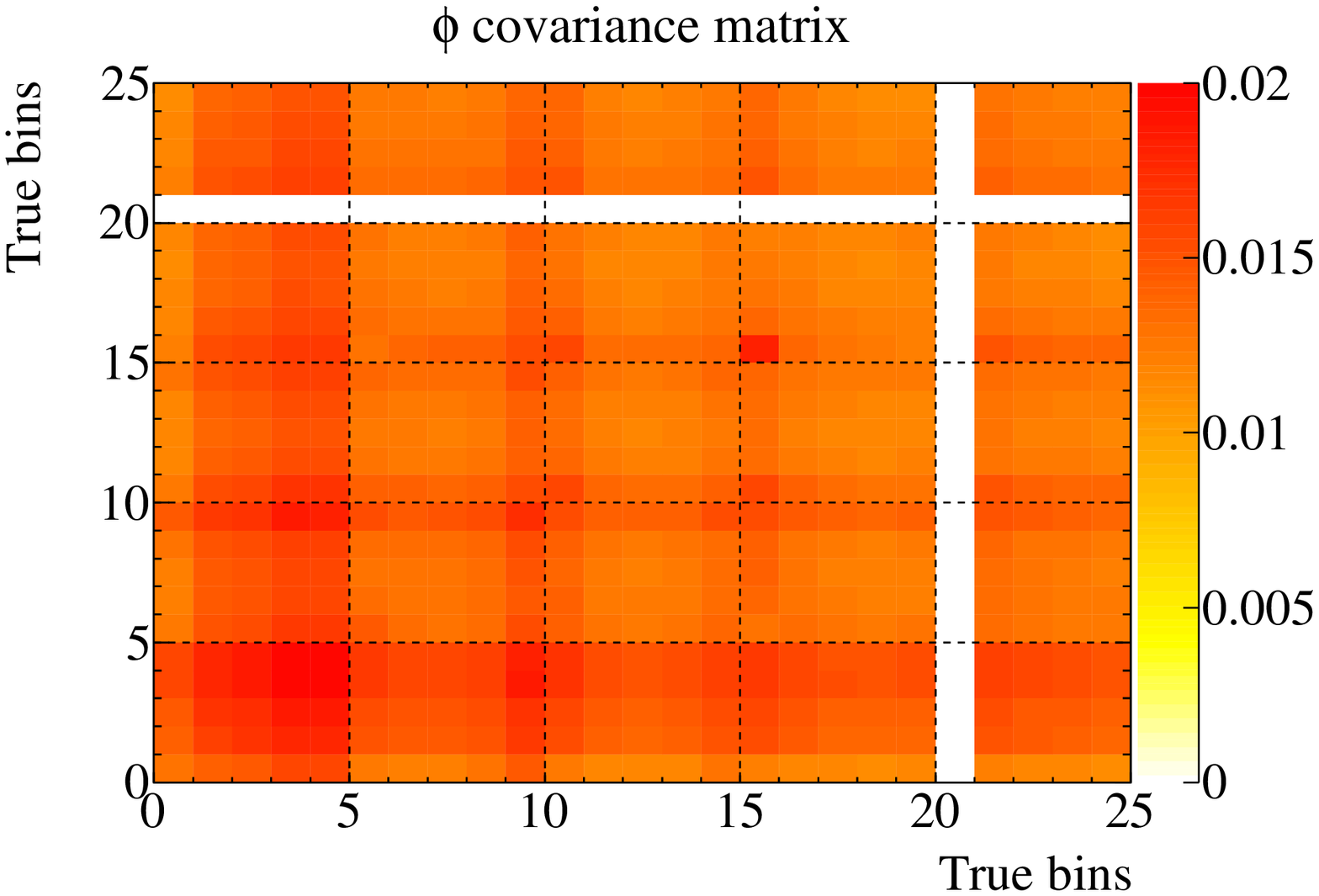}
  \includegraphics[width=1 \columnwidth]{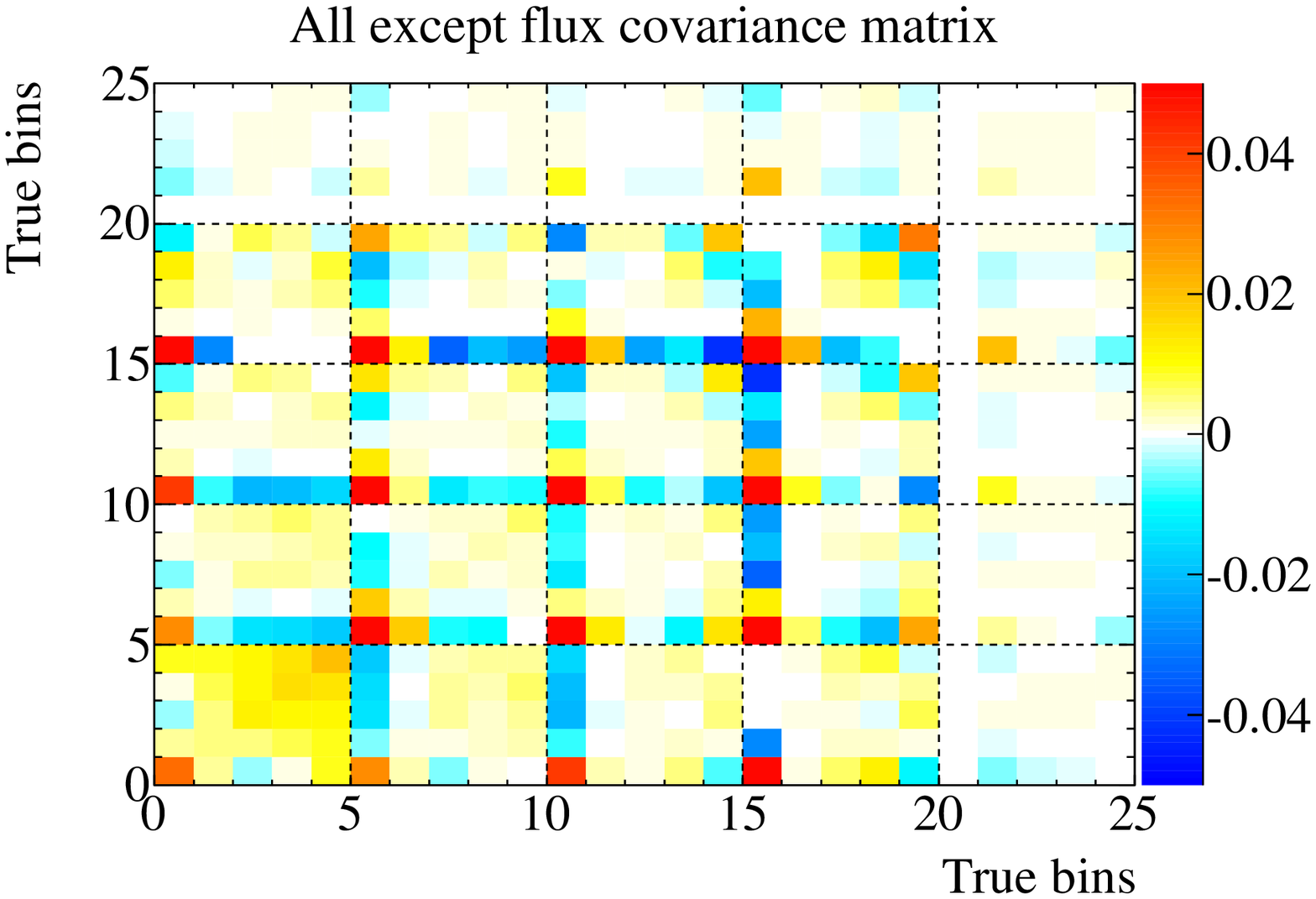}
  \caption{The fractional covariance matrix obtained from the uncertainties on the flux (top), and uncertainties due to all sources except the flux (bottom). The $x$-axis and $y$-axis bin numbering is defined in Table~\ref{tab:truebinning}. }
\label{fig:systcov}
\end{figure}

The propagation of the systematic error of the first four sources listed above is done by reweighting the NEUT MC. 
The correlations inside each source of systematic uncertainties are taken into account by generating a correlated set of systematic parameters that are used to reweight the MC. This procedure is repeated with 200 different sets of parameters (throws) and the rms of the difference between the cross-section result obtained for the nominal MC and the 200 throws is used to define a covariance matrix. The corresponding covariance matrix is given by
\begin{eqnarray}
V_{kl}=\frac{1}{200}\sum_{s=0}^{200}(\sigma_k^s-\sigma_k^{\rm{nom}})(\sigma_l^s-\sigma_l^{\rm{nom}}),
\end{eqnarray}
which is different for each source of systematic error and $\sigma^s_k$ is the cross section result obtained following Eq.~(\ref{eq:xs}) for the $s$th throw.
 The fractional covariance matrices, $V_{kl}/(\sigma_k^{\rm{nom}}\sigma_l^{\rm{nom}})$, due to the flux uncertainty and the rest of the sources are shown in Fig.~\ref{fig:systcov}.

\begin{table*}[!tpb]
 \caption{\label{tab:syserr} A summary of the systematic and statistical errors. The ``Unfold.,'' ``$\phi$,'' ``Det.,'' and ``FSI'' labels represent the systematic uncertainty due to the unfolding method, the beam flux, the detector response and the final state interactions, respectively, changed systematically following their respective covariance matrix. The ``Model'' column denotes the influence of changing all the neutrino interaction modeling parameters and channel rates. The ``Syst.,''``Stat.,'' and ``Tot.'' labels represent the systematic, statistic and total uncertainty respectively, where the error on the number of target nucleons (0.67\%) has been added in quadrature to the total systematic error. }
 \centering
 \begin{tabular}{c c| c c c c c |c|c||c }
 \hline
 \hline
$P_{\mu}$ (GeV/c) & 	 $\cos\theta_{\mu}$  &  Unfold. (\%)  & $\phi$(\%) &  Model (\%)  &  Det. (\%) &  FSI  (\%)  &  Syst.  (\%) &  Stat.  (\%) & Tot. (\%)  \\
 \hline
  $[0.0,0.4]$ 
 	 & $[-1,0]$ 	  & 	 0.5  & 	 11.4  & 	 18.0  & 	 2.1  & 	 0.5  & 	 21.4  & 	 2.0  & 	 21.5   \\ 
  	 & $[0,0.84]$ 	  & 	 0.6  & 	 12.8  & 	 5.5  & 	 3.6  & 	 1.2  & 	 14.5  & 	 4.9  & 	 15.3   \\ 
  	 & $[0.84,0.90]$ 	  & 	 0.2  & 	 13.1  & 	 10.8  & 	 2.7  & 	 1.4  & 	 17.3  & 	 9.5  & 	 19.7   \\ 
  	 & $[0.90,0.94]$ 	  & 	 1.0  & 	 14.1  & 	 10.7  & 	 5.0  & 	 3.6  & 	 18.8  & 	 12.3  & 	 22.4   \\ 
  	 & $[0.94,1]$ 	  & 	 0.3  & 	 14.0  & 	 12.9  & 	 4.9  & 	 3.0  & 	 20.0  & 	 14.7  & 	 24.8   \\ 
  \hline 
  $[0.4,0.5]$ 
 	 & $[-1,0]$ 	  & 	 1.2  & 	 12.0  & 	 39.5  & 	 2.7  & 	 0.9  & 	 41.4  & 	 3.2  & 	 41.5   \\ 
  	 & $[0,0.84]$ 	  & 	 0.2  & 	 11.4  & 	 5.7  & 	 1.3  & 	 0.3  & 	 12.8  & 	 4.2  & 	 13.5   \\ 
  	 & $[0.84,0.90]$ 	  & 	 0.6  & 	 11.4  & 	 5.0  & 	 1.0  & 	 0.4  & 	 12.5  & 	 8.6  & 	 15.2   \\ 
  	 & $[0.90,0.94]$ 	  & 	 0.5  & 	 11.7  & 	 5.4  & 	 1.3  & 	 0.5  & 	 13.0  & 	 10.1  & 	 16.4   \\ 
  	 & $[0.94,1]$ 	  & 	 0.5  & 	 13.1  & 	 7.2  & 	 2.3  & 	 0.9  & 	 15.2  & 	 11.7  & 	 19.2   \\ 
  \hline 
  $[0.5,0.7]$ 
 	 & $[-1,0]$ 	  & 	 1.2  & 	 12.6  & 	 46.1  & 	 1.9  & 	 0.4  & 	 47.9  & 	 9.0  & 	 48.7   \\ 
  	 & $[0,0.84]$ 	  & 	 0.3  & 	 11.1  & 	 3.8  & 	 1.1  & 	 0.4  & 	 11.8  & 	 3.8  & 	 12.4   \\ 
  	 & $[0.84,0.90]$ 	  & 	 0.3  & 	 10.8  & 	 3.4  & 	 0.8  & 	 0.3  & 	 11.4  & 	 6.2  & 	 13.0   \\ 
  	 & $[0.90,0.94]$ 	  & 	 0.4  & 	 11.0  & 	 5.7  & 	 0.8  & 	 0.4  & 	 12.5  & 	 7.3  & 	 14.4   \\ 
  	 & $[0.94,1]$ 	  & 	 0.0  & 	 11.6  & 	 11.4  & 	 1.1  & 	 0.3  & 	 16.4  & 	 7.9  & 	 18.2   \\ 
  \hline 
  $[0.7,0.9]$ 
 	 & $[-1,0]$ 	  & 	 1.8  & 	 13.5  & 	 148.3  & 	 2.0  & 	 0.6  & 	 149.0  & 	 31.5  & 	 152.3   \\ 
  	 & $[0,0.84]$ 	  & 	 0.4  & 	 11.4  & 	 3.2  & 	 1.1  & 	 0.4  & 	 11.9  & 	 5.1  & 	 12.9   \\ 
  	 & $[0.84,0.90]$ 	  & 	 0.4  & 	 10.9  & 	 5.9  & 	 0.8  & 	 0.2  & 	 12.5  & 	 6.8  & 	 14.2   \\ 
  	 & $[0.90,0.94]$ 	  & 	 0.2  & 	 10.7  & 	 11.1  & 	 1.1  & 	 0.5  & 	 15.5  & 	 7.7  & 	 17.3   \\ 
  	 & $[0.94,1]$ 	  & 	 0.2  & 	 11.0  & 	 17.6  & 	 0.9  & 	 0.4  & 	 20.8  & 	 7.0  & 	 21.9   \\ 
  \hline 
  $[0.9,30.0]$ 
 	 & $[-1,0]$ 	  & 	 -  & 	 -  & 	 -  & 	 -  & 	 -  & 	 - & 	 - & 	 - \\ 
	 & $[0,0.84]$ 	  & 	 0.2  & 	 11.9  & 	 5.6  & 	 1.4  & 	 0.6  & 	 13.3  & 	 5.4  & 	 14.3   \\ 
  	 & $[0.84,0.90]$ 	  & 	 0.2  & 	 11.3  & 	 2.5  & 	 0.9  & 	 0.3  & 	 11.7  & 	 5.8  & 	 13.1   \\ 
  	 & $[0.90,0.94]$ 	  & 	 0.4  & 	 11.1  & 	 2.3  & 	 0.7  & 	 0.4  & 	 11.4  & 	 5.2  & 	 12.5   \\ 
  	 & $[0.94,1]$ 	  & 	 0.2  & 	 10.9  & 	 2.3  & 	 0.8  & 	 0.3  & 	 11.2  & 	 2.9  & 	 11.6   \\ 
  \hline 

\hline
\end{tabular}
\end{table*}

The uncertainty due to the unfolding method, the fifth source of systematic error listed above, is estimated using the same fluctuated sample as for the statistical error. 
The difference between the nominal cross section and the average cross section extracted from these 1000 fluctuated sets is used as the estimate of the uncertainty due to the algorithm bias. The covariance matrix for this uncertainty is given by
\begin{eqnarray}
V_{kl}=(\langle \sigma_k \rangle-\sigma_k^{\rm{nom}})(\langle \sigma_l \rangle-\sigma_l^{\rm{nom}})
\end{eqnarray}
where $\langle \sigma_k \rangle$ is the mean of the distribution in bin $k$. The error due to the unfolding is generally below 1\% except for the backward bins.

The last uncertainty that needs to be taken into account is the one on the number of target nucleons, $T$, as explained in detail in Eq.~(\ref{eq:Tnucleon}) of Sec.~\ref{sec:nd280oa}. The resulting uncertainty of 0.67\% is then added in quadrature in all bins. Table~\ref{tab:syserr} summarizes the uncertainties for each bin.

The uncertainty on the cross section due to the flux is similar in each bin, and is dominated by the integrated flux normalization uncertainty of 10.9\%. 

 The errors due to the neutrino interaction modeling are very different from one bin to another.
Backward-going events contain the largest systematic uncertainty since we have a very low efficiency for this part of the phase space and the result is effectively a model-dependent extrapolation from the forward direction. The neutrino interaction modeling error is dominated by the uncertainty assigned to the spectral function corresponding to most of the systematic error at momenta between 700 MeV/c and 900 MeV/c and $\cos\theta_{\mu}>0.84$. This behavior is expected given the 30\% model-dependent difference in the muon energy spectrum shown in~\cite{Benhar:2005dj} and the way in which the unfolding process in these bins depends on it. For the other bins, the uncertainty on $M_A^{QE}$ is also important and dominant for momenta below 400 MeV/c.

The systematic error on the cross section due to the uncertainty on the detector response is bigger at low momentum, and is dominated by the external background uncertainty. For the other bins, the main contributions are shared more or less equally among the uncertainty in the magnetic field, FGD-TPC matching, charge confusion and the fiducial mass.

The systematic uncertainty due to final state interactions is relatively small, as its main effect is to change the number and energy of the pions, which are occasionally identified as the muon candidate. These constitute a small fraction of the sample and therefore the effect is small.

\section{Results}
\label{sec:results}
In this section, the flux-averaged double differential cross section is presented. In particular, we report the measurement for the forward bins in Sec.~\ref{sec:dxs}. In Sec.~\ref{sec:xstot}, we include the backward bin to give the flux-averaged total cross-section measurement.

\subsection{Flux-averaged differential cross section}
\label{sec:dxs}

\begin{table}[hbpt]
 \caption{\label{tab:ni}
The number of events in each bin.
Columns $3-5$ and $6-7$ give the information on the reconstructed and true variables, respectively. Columns $3-5$ give the number of selected events with NEUT, data and the number of background events respectively. 
 Columns $6-7$ give the number of simulated events of the NEUT MC and the number of inferred events for our data using the NEUT MC, respectively. The momentum bins are given in GeV/c. }
 \centering
 \begin{tabular}{ c c |c c c | c c }
 \hline
 \hline
 $P_{\mu}$ & 	 $\cos\theta_{\mu}$   & 	  $N_{j}^{\mathrm{neut}}$ & 	 $N_{j}^{\mathrm{data}}$ & 	$ B_{j}^{\mathrm{neut}}$ &    $N_{k}^{\mathrm{neut}}$ & 	 $\widehat{N}_{k}$ \\
 \hline
  $[0.0,0.4]$ 
 	 & $[-1,0]$ 	 & \multirow{2}{*}{ 555.3} & \multirow{2}{*}{ 556} & \multirow{2}{*}{ 142.8}    & 	 1149.0 & 	 1083.8\\ 
 	 & $[0,0.84]$ 	 &  	  & 	 & 	   & 	    	 1529.6 & 	 1521.3\\
 	 & $[0.84,0.90]$ 	 &   	  78.1 & 	 75 & 	 21.0  & 	 88.5 & 	 85.0 \\ 
 	 & $[0.90,0.94]$ 	 &   	  54.0 & 	 46 & 	 19.0  & 	 56.6 & 	 50.5 \\
 	 & $[0.94,1]$ 	 &   	  63.6 & 	 78 & 	 25.8  & 		 61.4 & 	 73.2 \\
  \hline 
  $[0.4,0.5]$ 
 	 & $[-1,0]$ 	 & \multirow{2}{*}{ 377.9} & \multirow{2}{*}{ 364} & \multirow{2}{*}{ 41.0}   & 	 70.7 & 	 69.1\\
 	 & $[0,0.84]$ 	 &  	  & 	 & 	   & 	   	 768.3 & 	 738.9\\
	 & $[0.84,0.90]$ 	 &   	  62.3 & 	 64 & 	 7.4  & 	 71.9 & 	 71.7\\
 	 & $[0.90,0.94]$ 	 &   	  43.8 & 	 45 & 	 6.2  & 	 44.2 & 	 42.3\\
 	 & $[0.94,1]$ 	 &   	  53.9 & 	 38 & 	 12.8  & 	 	 50.1 & 	 38.7\\
  \hline 
  $[0.5,0.7]$ 
 	 & $[-1,0]$ 	 & \multirow{2}{*}{ 497.8} & \multirow{2}{*}{ 475} & \multirow{2}{*}{ 48.9}    & 	 12.3 & 	 11.1\\
 	 & $[0,0.84]$ 	 &  	  & 	 & 	   & 	    	 865 & 	 820.2\\
 	 & $[0.84,0.90]$ 	 &   	  138.1 & 	 133 & 	 11.2  & 	 	 175.3 & 	 163.4\\
 	 & $[0.90,0.94]$ 	 &   	  98.3 & 	 81 & 	 9.9  & 	 112.9 & 	 95.0 \\
 	 & $[0.94,1]$ 	 &   	  130.8 & 	 122 & 	 26.3  &  	 126.0 & 	 113.1 \\
  \hline 
  $[0.7,0.9]$ 
 	 & $[-1,0]$ 	 & \multirow{2}{*}{ 211.4} & \multirow{2}{*}{ 198} & \multirow{2}{*}{ 23.7}  & 0.6 & 	 0.6\\
 	 & $[0,0.84]$ 	 &  	  & 	 & 	   & 	    	 287.1 & 	 267.1\\
 	 & $[0.84,0.90]$ 	 &   	  94.5 & 	 74 & 	 8.5  & 	 110.3 & 	 91.4\\
 	 & $[0.90,0.94]$ 	 &   	  73.5 & 	 57 & 	 5.1  & 	 	 80.3 & 	 64.0\\
 	 & $[0.94,1]$ 	 &   	  111.5 & 	 105 & 	 13.9  & 	 106.0 & 	 98.2\\
  \hline 
  $[0.9,30]$ 
 	 & $[-1,0]$ 	 & \multirow{2}{*}{ 301.6} & \multirow{2}{*}{ 282} & \multirow{2}{*}{ 37.8}  & 	 0.0 & 	 0.0 \\
 	 & $[0,0.84]$ 	 &  	  & 	 & 	   & 	    	 335 & 	 310.7\\
 	 & $[0.84,0.90]$ 	 &   	  242.6 & 	 219 & 	 24.6  & 	 287.8 & 	 256.8 \\
 	 & $[0.90,0.94]$ 	 &   	  294.0 & 	 262 & 	 24.2  & 	 350.6 & 	 309.7 \\
 	 & $[0.94,1]$ 	 &   	  1240.7 & 	 1211 & 	 113.9  &  	 1536.6 & 	 1488.6 \\
  \hline 
  &  Total  &   	 4723.5  & 	 4485 & 	 624.0  & 	 8276.2 & 	 7864.5  \\

    \hline
    \hline
  \end{tabular}
\end{table}

The flux-averaged double differential cross section is calculated as,
\begin{eqnarray}
\left\langle \frac{\partial \sigma}{\partial p_{\mu}\partial{\cos \theta_{\mu}}}\right\rangle_{\alpha\beta}=\frac{\widehat{N}_{\alpha\beta}}{T\phi \Delta p_{\mu,\alpha} \Delta \cos\theta_{\mu,\beta}}
\end{eqnarray}
 where $\widehat{N}_{\alpha\beta}$ is the number of inferred events for the two-dimensional binning in momentum and angle labeled by $\alpha$ and $\beta$, respectively, while $\Delta p_{\mu,\alpha}$ and $\Delta \cos\theta_{\mu,\beta}$ give the respective bin widths.
Table~\ref{tab:ni} gives the number of reconstructed events in data and the background prediction needed to calculate the number of inferred events as defined in Eq.~(\ref{eq:xs}). The numbers of simulated and predicted CC events in the FGD volume using the NEUT generator are also shown together with the result of the unfolding.

\begin{figure*}[!hbpt]
  \centering
  \includegraphics[width=0.4\textwidth]{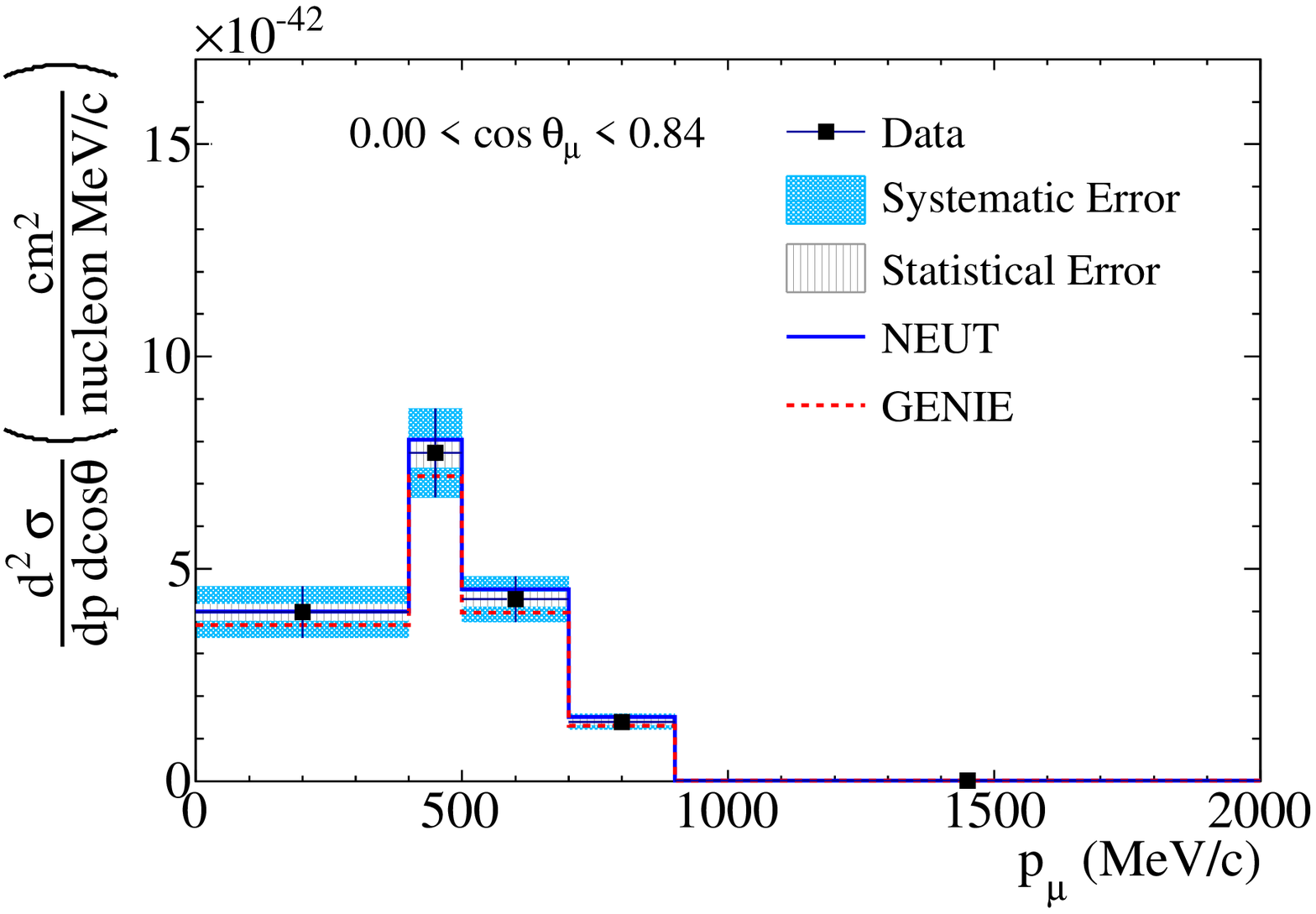}
  \includegraphics[width=0.4\textwidth]{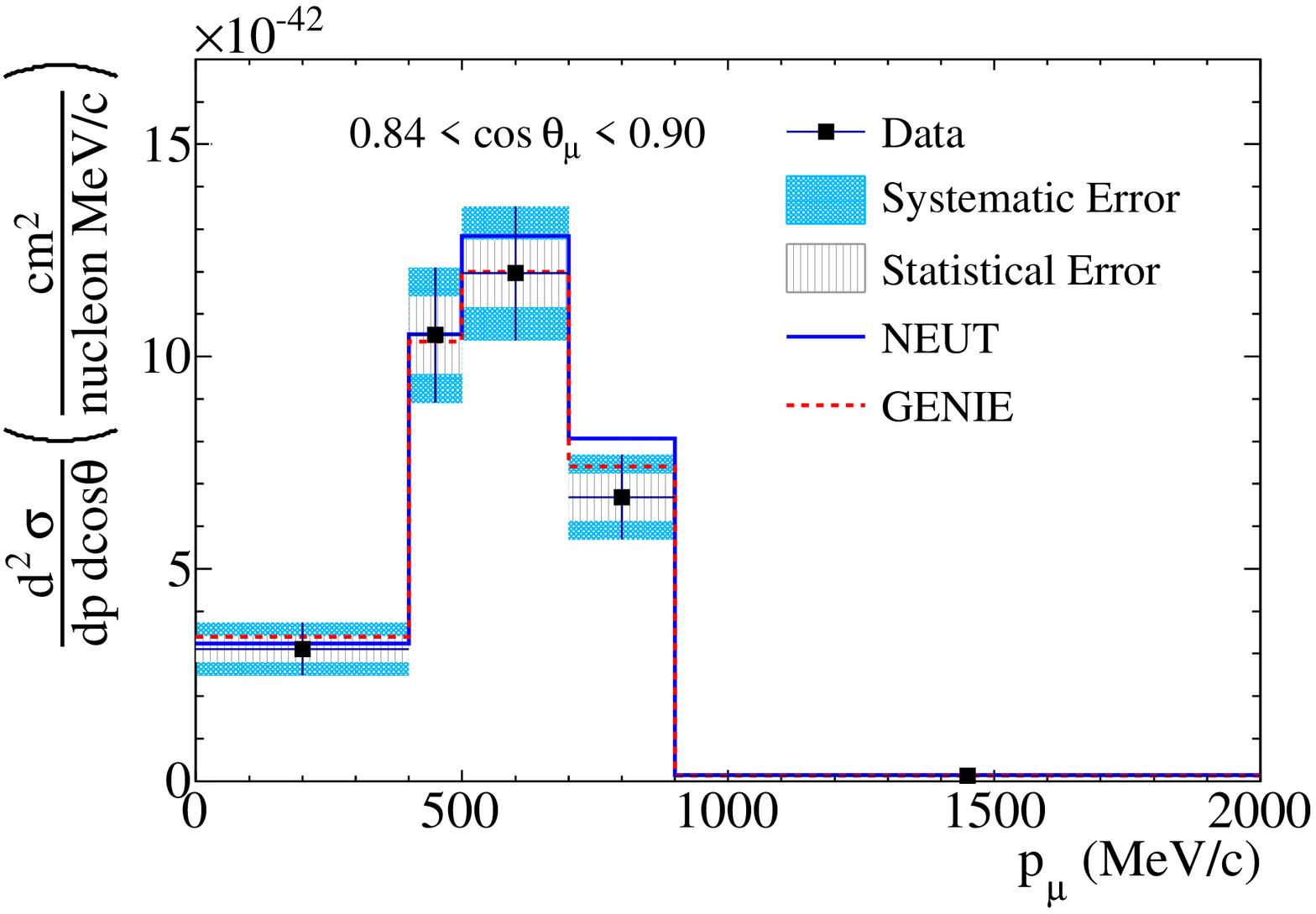}
  \includegraphics[width=0.4\textwidth]{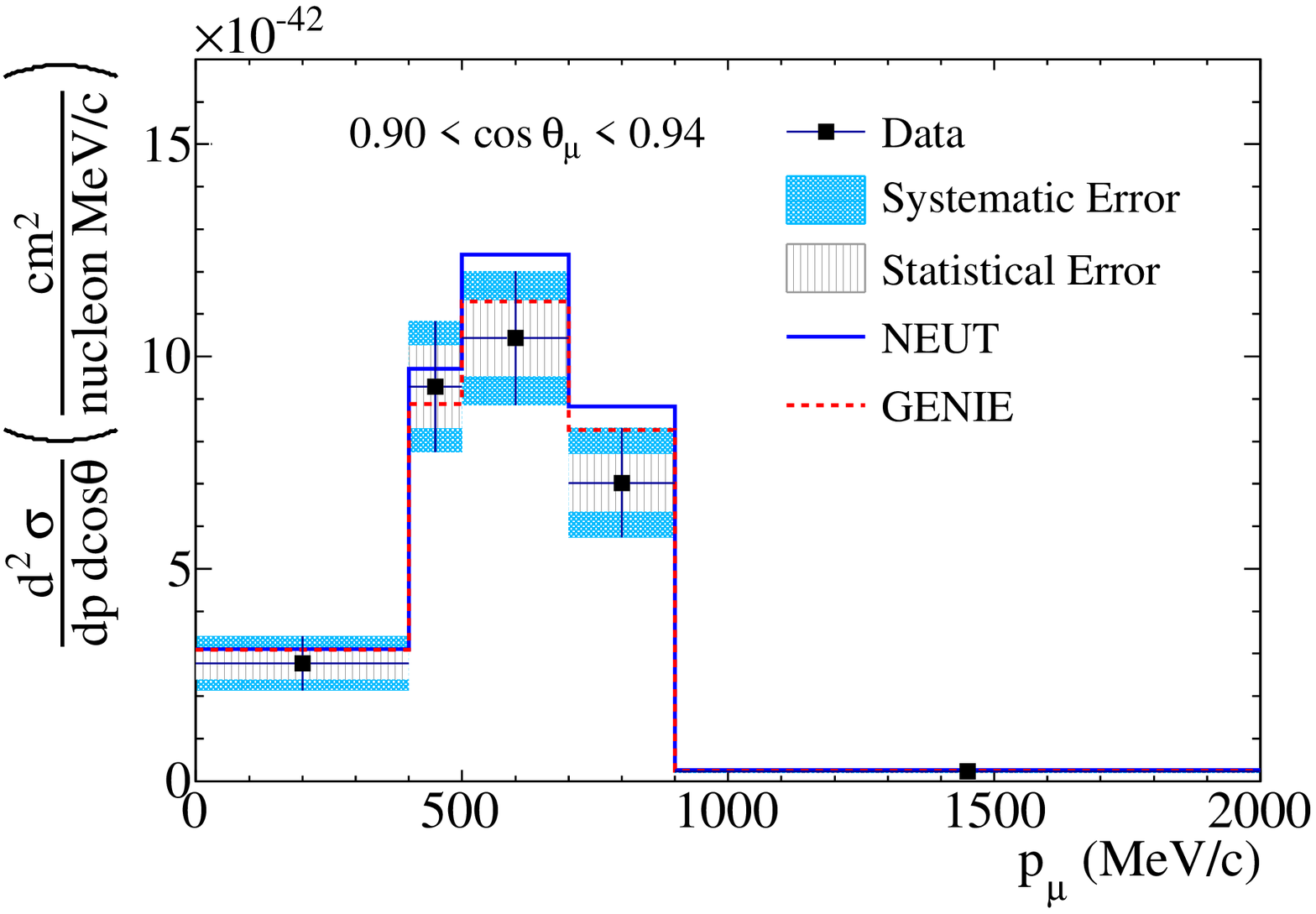}
  \includegraphics[width=0.4\textwidth]{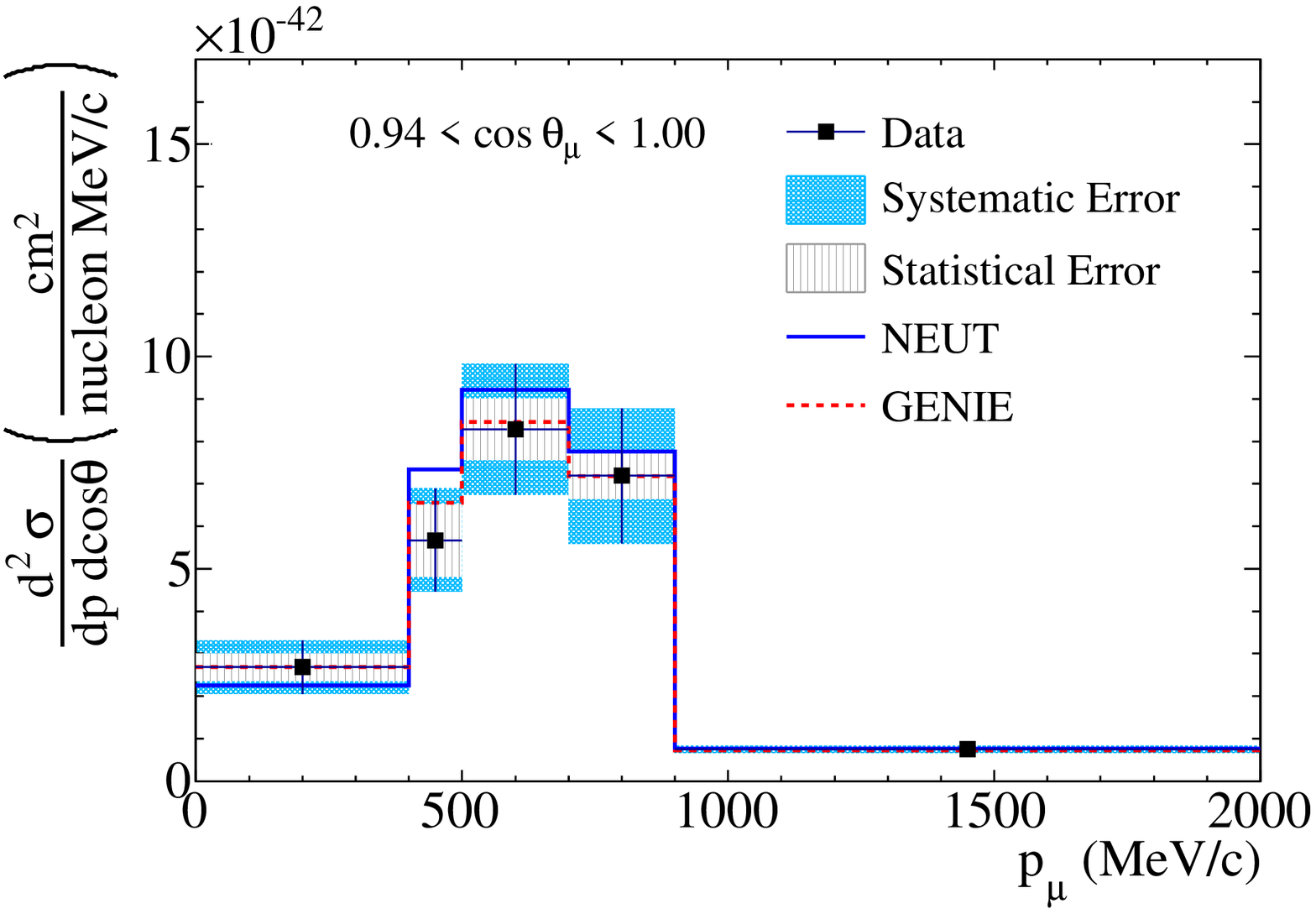}
 \caption{The CC-inclusive differential cross section in cm$^2$/nucleon/MeV, with statistical and systematic errors. Each graph corresponds to a bin in muon angle. }
\label{fig:dxs}
\end{figure*}

The measured differential cross section is shown in Table~\ref{tab:result} and Fig.~\ref{fig:dxs} only for the forward bins since the results obtained for the backward bins are completely model dependent, while the total fractional covariance matrix is given in Table~\ref{tab:totcov}.
In the case of the backward bins, we rely entirely on the MC to determine the fraction of the selected forward-going events which are due to true backward-going events.

The cross section was also calculated unfolding with the GENIE generator.
The results are consistent within the errors estimated from modeling uncertainties. 

\begin{table}[hbpt]
\centering
\caption{\label{tab:result} The differential cross-section measurement, with its statistical and systematic errors (labelled by ``Stat. err.'' and ``Syst. err.'' respectively) where the number of target nucleons is included into the total systematic error listed here.}
 \centering
 \begin{tabular}{c c| c c c c }
 \hline
 \hline
$P_{\mu}$  & 	 $\cos\theta_{\mu}$  & 	  $ \langle \frac{\partial^2\sigma}{\partial p_{\mu} \partial \cos \theta_{\mu}} \rangle $ 	 & Stat. err.	 & Syst. err. \\
 GeV/c & 	 & 	  $\rm{(cm^2/\mathrm{nucleon}/MeV)}$ 	 & \% 	 &  \%  \\
\hline

  $[0.0,0.4]$ 
	 & $[0,0.84]$ 	  & 	 3.98 $\times 10^{-42}$            & 	 5.0  & 	 14.5 \\ 
	 & $[0.84,0.90]$ 	  & 	 3.11 $\times 10^{-42}$    & 	 9.5  & 	 17.3 \\ 
	 & $[0.90,0.94]$ 	  & 	 2.77 $\times 10^{-42}$    & 	 12.3  & 	 18.8 \\ 
	 & $[0.94,1]$ 	  & 	 2.68 $\times 10^{-42}$            & 	 14.7  & 	 20.0 \\ 
\hline 
  $[0.4,0.5]$ 
	 & $[0,0.84]$ 	  & 	 7.73 $\times 10^{-42}$             & 	 4.2  & 	 12.8 \\ 
	 & $[0.84,0.90]$ 	  & 	 10.50 $\times 10^{-42}$    & 	 8.6  & 	 12.5 \\ 
	 & $[0.90,0.94]$ 	  & 	 9.29 $\times 10^{-42}$    & 	 10.1  & 	 13.0 \\ 
	 & $[0.94,1]$ 	  & 	 5.67 $\times 10^{-42}$            & 	 11.7  & 	 15.2 \\ 
\hline 
  $[0.5,0.7]$ 
	 & $[0,0.84]$ 	  & 	 4.29 $\times 10^{-42}$             & 	 3.8  & 	 11.8 \\ 
	 & $[0.84,0.90]$ 	  & 	 11.96 $\times 10^{-42}$    & 	 6.2  & 	 11.5 \\ 
	 & $[0.90,0.94]$ 	  & 	 10.43 $\times 10^{-42}$    & 	 7.3  & 	 12.5 \\ 
	 & $[0.94,1]$ 	  & 	 8.28 $\times 10^{-42}$             & 	 7.9  & 	 16.4 \\ 
\hline 
  $[0.7,0.9]$ 
	 & $[0,0.84]$ 	  & 	 1.40 $\times 10^{-42}$            & 	 5.1  & 	 11.9 \\ 
	 & $[0.84,0.90]$ 	  & 	 6.69 $\times 10^{-42}$    & 	 6.8  & 	 12.5 \\ 
	 & $[0.90,0.94]$ 	  & 	 7.03 $\times 10^{-42}$    & 	 7.7  & 	 15.5 \\ 
	 & $[0.94,1]$ 	  & 	 7.19 $\times 10^{-42}$            & 	 7.0  & 	 20.8 \\ 
\hline 
  $[0.9,30.0]$ 
	 & $[0,0.84]$ 	  & 	 0.01 $\times 10^{-42}$            & 	 5.4  & 	 13.3 \\ 
	 & $[0.84,0.90]$ 	  & 	 0.13 $\times 10^{-42}$    & 	 5.9  & 	 11.7 \\ 
	 & $[0.90,0.94]$ 	  & 	 0.23 $\times 10^{-42}$    & 	 5.2  & 	 11.4 \\ 
	 & $[0.94,1]$ 	  & 	 0.75 $\times 10^{-42}$            & 	 2.9  & 	 11.2 \\ 
\hline 

\hline
\hline
\end{tabular}
\end{table}

\begin{figure}[!h]
  \centering
  \includegraphics[width=0.5\textwidth]{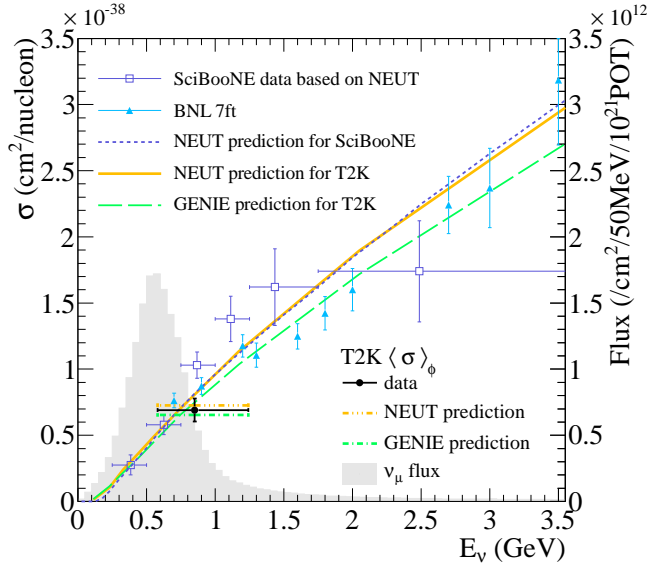}
  \caption{The T2K total flux-averaged cross section with the NEUT and the GENIE prediction for T2K and SciBooNE. The T2K data point is placed at the flux mean energy. The vertical error represents the total (statistical and systematic) uncertainty, and the horizontal bar represent 68\% of the flux at each side of the mean energy.
The T2K flux distribution is shown in grey. The predictions for SciBooNE have been done for a C$_8$H$_8$ target \cite{Nakajima:2010fp} which is comparable to the mixed T2K target. BNL data has been measured on deuterium \cite{Mukhin:1979bd}.}
\label{fig:res}
\end{figure}

\begin{table*}[!hb]
 \caption{\label{tab:totcov}
The total fractional covariance matrix including statistical and systematic errors. The values are given in (\%). The bin numbering is defined in Table~\ref{tab:truebinning}.}
 \centering
 \begin{tabular}{ c |c c c c c | c c c c c |c c c c c |c c c c c |c c c c c  }
 \hline
 \hline
 Bin \#	 & 0 	 & 1 	 & 2 	 & 3 	 & 4 	 & 5 	 & 6 	 & 7 	 & 8 	 & 9 	 & 10 	 & 11 	 & 12 	 & 13 	 & 14 	 & 15 	 & 16 	 & 17 	 & 18 	 & 19 	 & 20 	 & 21 	 & 22 	 & 23 	 & 24 	 \\ 
\hline 
 24 	 & 1.2  	& 1.4  	& 1.5  	& 1.6  	& 1.6  	& 0.8  	& 1.2  	& 1.3  	& 1.3  	& 1.5  	& 1.3  	& 1.2  	& 1.2  	& 1.3  	& 1.2  	& 0.9  	& 1.3  	& 1.3  	& 1.3  	& 1.0  	& 0.0  	& 1.3  	& 1.3  	& 1.3  	& 1.4  	 \\ 
23 	 & 1.1  	& 1.4  	& 1.5  	& 1.6  	& 1.6  	& 1.3  	& 1.3  	& 1.3  	& 1.3  	& 1.5  	& 1.4  	& 1.3  	& 1.2  	& 1.2  	& 1.4  	& 1.3  	& 1.3  	& 1.2  	& 1.2  	& 1.3  	& 0.0  	& 1.4  	& 1.4  	& 1.6  	& 1.3  	 \\ 
22 	 & 1.0  	& 1.5  	& 1.6  	& 1.6  	& 1.6  	& 1.4  	& 1.3  	& 1.3  	& 1.3  	& 1.5  	& 1.5  	& 1.3  	& 1.3  	& 1.2  	& 1.4  	& 1.5  	& 1.3  	& 1.3  	& 1.1  	& 1.3  	& 0.0  	& 1.5  	& 1.8  	& 1.4  	& 1.3  	 \\ 
21 	 & 0.8  	& 1.4  	& 1.6  	& 1.6  	& 1.4  	& 1.8  	& 1.4  	& 1.4  	& 1.3  	& 1.6  	& 2.4  	& 1.4  	& 1.2  	& 1.2  	& 1.4  	& 3.6  	& 1.6  	& 1.1  	& 1.0  	& 1.4  	& 0.0  	& 2.1  	& 1.5  	& 1.4  	& 1.3  	 \\ 
20 	 & 0.0  	& 0.0  	& 0.0  	& 0.0  	& 0.0  	& 0.0  	& 0.0  	& 0.0  	& 0.0  	& 0.0  	& 0.0  	& 0.0  	& 0.0  	& 0.0  	& 0.0  	& 0.0  	& 0.0  	& 0.0  	& 0.0  	& 0.0  	& 0.0  	& 0.0  	& 0.0  	& 0.0  	& 0.0  	 \\ 
\hline 
 19 	 & 0.1  	& 1.5  	& 2.1  	& 1.9  	& 1.3  	& 3.9  	& 1.8  	& 1.6  	& 1.1  	& 2.0  	& -1.4  	& 1.5  	& 1.5  	& 0.6  	& 3.4  	& -4.1  	& 1.2  	& 0.7  	& -0.2  	& 4.9  	& 0.0  	& 1.4  	& 1.3  	& 1.3  	& 1.0  	 \\ 
18 	 & 2.4  	& 1.6  	& 1.3  	& 1.8  	& 2.3  	& -0.7  	& 0.9  	& 1.1  	& 1.6  	& 1.4  	& 1.4  	& 1.1  	& 1.2  	& 2.0  	& 0.4  	& 0.5  	& 1.2  	& 1.9  	& 3.3  	& -0.2  	& 0.0  	& 1.0  	& 1.1  	& 1.2  	& 1.3  	 \\ 
17 	 & 1.8  	& 1.6  	& 1.5  	& 1.8  	& 2.0  	& 0.4  	& 1.2  	& 1.3  	& 1.4  	& 1.5  	& 0.9  	& 1.2  	& 1.5  	& 1.5  	& 1.0  	& -0.7  	& 1.3  	& 2.2  	& 1.9  	& 0.7  	& 0.0  	& 1.1  	& 1.3  	& 1.2  	& 1.3  	 \\ 
16 	 & 1.3  	& 1.5  	& 1.6  	& 1.6  	& 1.6  	& 1.9  	& 1.3  	& 1.3  	& 1.3  	& 1.5  	& 2.3  	& 1.4  	& 1.2  	& 1.2  	& 1.3  	& 3.6  	& 1.7  	& 1.3  	& 1.2  	& 1.2  	& 0.0  	& 1.6  	& 1.3  	& 1.3  	& 1.3  	 \\ 
15 	 & 11.0  	& -1.3  	& -3.8  	& -4.2  	& -4.0  	& 35.8  	& 2.6  	& -2.0  	& -0.6  	& -1.0  	& 59.1  	& 3.3  	& -1.1  	& 0.1  	& -2.7  	& 222.4  	& 3.6  	& -0.7  	& 0.5  	& -4.1  	& 0.0  	& 3.6  	& 1.5  	& 1.3  	& 0.9  	 \\ 
\hline 
 14 	 & 0.6  	& 1.6  	& 2.0  	& 2.0  	& 1.7  	& 2.9  	& 1.7  	& 1.6  	& 1.4  	& 2.3  	& -0.2  	& 1.5  	& 1.5  	& 1.0  	& 3.5  	& -2.7  	& 1.3  	& 1.0  	& 0.4  	& 3.4  	& 0.0  	& 1.4  	& 1.4  	& 1.4  	& 1.2  	 \\ 
13 	 & 1.7  	& 1.6  	& 1.5  	& 1.8  	& 2.0  	& 0.2  	& 1.1  	& 1.3  	& 1.8  	& 1.6  	& 1.1  	& 1.2  	& 1.4  	& 2.3  	& 1.0  	& 0.1  	& 1.2  	& 1.5  	& 2.0  	& 0.6  	& 0.0  	& 1.2  	& 1.2  	& 1.2  	& 1.3  	 \\ 
12 	 & 1.3  	& 1.5  	& 1.6  	& 1.7  	& 1.8  	& 1.3  	& 1.3  	& 1.4  	& 1.4  	& 1.6  	& 0.5  	& 1.3  	& 1.7  	& 1.4  	& 1.5  	& -1.1  	& 1.2  	& 1.5  	& 1.2  	& 1.5  	& 0.0  	& 1.2  	& 1.3  	& 1.2  	& 1.2  	 \\ 
11 	 & 1.5  	& 1.5  	& 1.4  	& 1.5  	& 1.5  	& 2.6  	& 1.5  	& 1.2  	& 1.2  	& 1.5  	& 2.1  	& 1.6  	& 1.3  	& 1.2  	& 1.5  	& 3.3  	& 1.4  	& 1.2  	& 1.1  	& 1.5  	& 0.0  	& 1.4  	& 1.3  	& 1.3  	& 1.2  	 \\ 
10 	 & 5.4  	& 0.7  	& -0.5  	& -0.3  	& 0.1  	& 13.8  	& 1.9  	& 0.1  	& 0.6  	& 1.0  	& 23.1  	& 2.1  	& 0.5  	& 1.1  	& -0.2  	& 59.1  	& 2.3  	& 0.9  	& 1.4  	& -1.4  	& 0.0  	& 2.4  	& 1.5  	& 1.4  	& 1.3  	 \\ 
\hline 
 9 	 & 1.6  	& 1.9  	& 2.1  	& 2.5  	& 2.7  	& 1.5  	& 1.6  	& 1.7  	& 2.1  	& 4.6  	& 1.0  	& 1.5  	& 1.6  	& 1.6  	& 2.3  	& -1.0  	& 1.5  	& 1.5  	& 1.4  	& 2.0  	& 0.0  	& 1.6  	& 1.5  	& 1.5  	& 1.5  	 \\ 
8 	 & 1.5  	& 1.7  	& 1.8  	& 2.3  	& 2.1  	& 0.4  	& 1.3  	& 1.6  	& 2.8  	& 2.1  	& 0.6  	& 1.2  	& 1.4  	& 1.8  	& 1.4  	& -0.6  	& 1.3  	& 1.4  	& 1.6  	& 1.1  	& 0.0  	& 1.3  	& 1.3  	& 1.3  	& 1.3  	 \\ 
7 	 & 0.8  	& 1.6  	& 2.1  	& 2.1  	& 1.9  	& 0.4  	& 1.3  	& 2.3  	& 1.6  	& 1.7  	& 0.1  	& 1.2  	& 1.4  	& 1.3  	& 1.6  	& -2.0  	& 1.3  	& 1.3  	& 1.1  	& 1.6  	& 0.0  	& 1.4  	& 1.3  	& 1.3  	& 1.3  	 \\ 
6 	 & 1.6  	& 1.6  	& 1.4  	& 1.6  	& 1.5  	& 3.1  	& 1.8  	& 1.3  	& 1.3  	& 1.6  	& 1.9  	& 1.5  	& 1.3  	& 1.1  	& 1.7  	& 2.6  	& 1.3  	& 1.2  	& 0.9  	& 1.8  	& 0.0  	& 1.4  	& 1.3  	& 1.3  	& 1.2  	 \\ 
5 	 & 4.1  	& 1.1  	& 0.2  	& 0.3  	& 0.1  	& 17.2  	& 3.1  	& 0.4  	& 0.4  	& 1.5  	& 13.8  	& 2.6  	& 1.3  	& 0.2  	& 2.9  	& 35.8  	& 1.9  	& 0.4  	& -0.7  	& 3.9  	& 0.0  	& 1.8  	& 1.4  	& 1.3  	& 0.8  	 \\ 
\hline 
 4 	 & 2.6  	& 2.7  	& 3.1  	& 4.0  	& 5.5  	& 0.1  	& 1.5  	& 1.9  	& 2.1  	& 2.7  	& 0.1  	& 1.5  	& 1.8  	& 2.0  	& 1.7  	& -4.0  	& 1.6  	& 2.0  	& 2.3  	& 1.3  	& 0.0  	& 1.4  	& 1.6  	& 1.6  	& 1.6  	 \\ 
3 	 & 1.7  	& 2.5  	& 3.4  	& 5.4  	& 4.0  	& 0.3  	& 1.6  	& 2.1  	& 2.3  	& 2.5  	& -0.3  	& 1.5  	& 1.7  	& 1.8  	& 2.0  	& -4.2  	& 1.6  	& 1.8  	& 1.8  	& 1.9  	& 0.0  	& 1.6  	& 1.6  	& 1.6  	& 1.6  	 \\ 
2 	 & 1.1  	& 2.3  	& 4.0  	& 3.4  	& 3.1  	& 0.2  	& 1.4  	& 2.1  	& 1.8  	& 2.1  	& -0.5  	& 1.4  	& 1.6  	& 1.5  	& 2.0  	& -3.8  	& 1.6  	& 1.5  	& 1.3  	& 2.1  	& 0.0  	& 1.6  	& 1.6  	& 1.5  	& 1.5  	 \\ 
1 	 & 1.8  	& 2.3  	& 2.3  	& 2.5  	& 2.7  	& 1.1  	& 1.6  	& 1.6  	& 1.7  	& 1.9  	& 0.7  	& 1.5  	& 1.5  	& 1.6  	& 1.6  	& -1.3  	& 1.5  	& 1.6  	& 1.6  	& 1.5  	& 0.0  	& 1.4  	& 1.5  	& 1.4  	& 1.4  	 \\ 
0 	 & 4.6  	& 1.8  	& 1.1  	& 1.7  	& 2.6  	& 4.1  	& 1.6  	& 0.8  	& 1.5  	& 1.6  	& 5.4  	& 1.5  	& 1.3  	& 1.7  	& 0.6  	& 11.0  	& 1.3  	& 1.8  	& 2.4  	& 0.1  	& 0.0  	& 0.8  	& 1.0  	& 1.1  	& 1.2  	 \\ 
\hline 
 
    \hline
    \hline
  \end{tabular}
\end{table*}

\subsection{Flux-averaged total cross section}
\label{sec:xstot}

The flux-averaged total cross section is calculated by taking all the bins into account, including the backward bins, although this analysis has very low efficiency in the backward bins resulting in larger statistical errors and model dependence.

 The flux-averaged total cross section is calculated as,
\begin{eqnarray}
 \langle \sigma \rangle_{\phi}= \frac{ \widehat{N}^{\rm{tot}}}{T\phi}
\end{eqnarray}
where $\widehat{N}^{\rm{tot}}$ is given in Table~\ref{tab:ni}.
 We obtain
\begin{eqnarray}
\langle \sigma_{ \rm{CC}} \rangle_{\phi} &=&(6.91 \pm 0.13 (stat) \pm 0.84 (syst))  \nonumber \\
& &\times 10^{-39} \rm{\frac{cm^2}{nucleon}}
\end{eqnarray}
which agrees well with the MC predicted values from NEUT and GENIE that are
\begin{eqnarray}
\langle \sigma_{ \rm{CC}}^{\rm{NEUT}} \rangle_{\phi}&=&7.27 \times 10^{-39}  \rm{\frac{cm^2}{nucleon}}  \\
\langle \sigma_{ \rm{CC}}^{\rm{GENIE}} \rangle_{\phi} &=&6.54 \times 10^{-39}  \rm{\frac{cm^2}{nucleon}}
\end{eqnarray}
The total cross section result for T2K is shown in Fig.~\ref{fig:res} together with CC-inclusive measurements from other experiments.

\subsection{Future improvements}
There is ongoing work in T2K expected to reduce the model dependence in
the neutrino event generator and to reduce the flux errors. These studies
should lead to improved CC-inclusive cross-section results in the future.

\section{Summary}
\label{sec:conclusion}
We have selected a sample of $\nu_{\mu}$ CC-inclusive interactions in the tracker of the T2K off-axis near detector. 
We used a largely model-independent method to extract the flux-averaged differential $\nu_{\mu}$ CC cross section in muon momentum and angle.
These results represent the first such measurement done for CC-inclusive interactions on carbon at a mean neutrino energy of 0.85~GeV and are presented as a two-dimensional differential cross section. In addition, these results have been used to calculate a flux-averaged total CC-inclusive cross section that is consistent with previous measurements in this energy range. The data related to this measurement can be found electronically in \cite{t2kdata}.

\begin{acknowledgments}
We thank the J-PARC accelerator team for the superb accelerator performance and
CERN NA61 colleagues for providing essential particle production data and for their fruitful collaboration.
We acknowledge the support of MEXT, Japan; 
NSERC, NRC and CFI, Canada;
CEA and CNRS/IN2P3, France;
DFG, Germany; 
INFN, Italy;
Ministry of Science and Higher Education, Poland;
RAS, RFBR and the Ministry of Education and Science
of the Russian Federation; 
MEST and NRF, South Korea;
MICINN and CPAN, Spain;
SNSF and SER, Switzerland;
STFC, UK; NSF and 
DOE, USA
We also thank CERN for their donation of the UA1/NOMAD magnet 
and DESY for the HERA-B magnet mover system.
In addition, participation of individual researchers
and institutions in T2K has been further supported by funds from: ERC (FP7), EU; JSPS, Japan; Royal Society, UK; 
DOE Early Career program, and the A. P. Sloan Foundation, USA.
\end{acknowledgments}

\pagebreak
\pagebreak
\pagebreak
\newpage

\bibliography{cc_xs_paper}

\end{document}